\documentclass[11pt, leqno]{article}
\usepackage{xcolor,soul}
\usepackage[pdftex,colorlinks=true]{hyperref}
\definecolor{darkblue}{rgb}{0,0,.6}
\hypersetup{citecolor=darkblue,linkcolor=darkblue,urlcolor=darkblue}
\usepackage{amsmath,setspace,multirow,lineno}
\usepackage{orcidlink}
\usepackage[font=small,labelfont=bf]{caption}
\usepackage[top=1.05in, bottom=1.05in, left=1in, right=1in]{geometry}

\usepackage{natbib}

\DeclareCaptionStyle{italic}[justification=centering]{labelfont={bf},textfont={it},labelsep=colon}
\usepackage{graphicx,psfrag,epsf,textcomp,epstopdf, amsthm, paralist}
\usepackage{enumerate, dsfont, alltt, verbatim}
\usepackage{natbib}
\usepackage{url,xcolor}
\usepackage{booktabs, subfig, bm, paralist,mathpazo,tikz,todonotes,longtable,microtype}
\usepackage[linesnumbered,ruled,vlined]{algorithm2e}

\definecolor{DarkRed}{rgb}{.7,0,.4}

\newcommand{\argmax}{\operatornamewithlimits{argmax}}

\newcommand{\blind}{0}
\DeclareMathOperator{\atantwo}{atan2}

\newcommand{\X}{\mathcal{X}}
\newcommand{\Y}{\mathcal{Y}}

\newcommand{\Rlogo}{\protect\includegraphics[height=1.8ex,keepaspectratio]{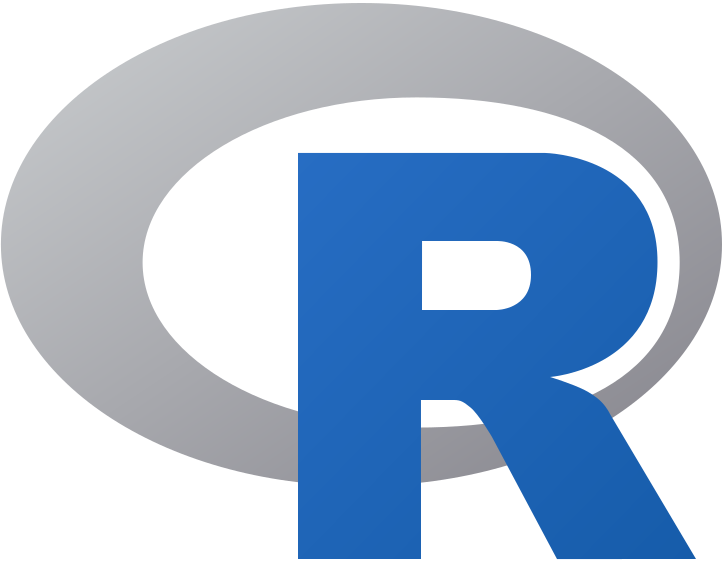}}

\graphicspath{{plots/}}
\DeclareMathOperator*{\argmin}{\arg\!\min}

\newsavebox\CBox

 \newtheorem{@definition}{\sc Definition}[section]

 \newtheorem{corollary}{\sc Corollary}[section]
 \newtheorem{theorem}{\sc Theorem}[section]

  \renewcommand\X{\mathcal{X}}

\date{}

\begin{document}

\def\spacingset#1{\renewcommand{\baselinestretch}{#1}\small\normalsize} \spacingset{1}

\if0\blind
{
\title{\bf Enhancing Spatial Functional Linear Regression with Robust Dimension Reduction Methods}}
\author{
Ufuk Beyaztas\footnote{Postal address: Department of Statistics, Marmara University, 34722, Kadikoy-Istanbul, Turkey; Email: ufuk.beyaztas@marmara.edu.tr} \orcidlink{0000-0002-5208-4950} 
\\
Department of Statistics \\
Marmara University \\
\\
Abhijit Mandal \orcidlink{0000-0003-3830-6526} 
\\
    Department of Mathematical Sciences \\
     	University of Texas at El Paso \\
\\
Han Lin Shang \orcidlink{0000-0003-1769-6430} 
\\
    Department of Actuarial Studies and Business Analytics \\
    Macquarie University
}
\maketitle
\fi

\if1\blind
{
\title{\bf Enhancing Spatial Functional Linear Regression with Robust Dimension Reduction Methods}
} \fi

\maketitle

\begin{abstract}
This paper introduces a robust estimation strategy for the spatial functional linear regression model using dimension reduction methods, specifically functional principal component analysis (FPCA) and functional partial least squares (FPLS). These techniques are designed to address challenges associated with spatially correlated functional data, particularly the impact of outliers on parameter estimation. By projecting the infinite-dimensional functional predictor onto a finite-dimensional space defined by orthonormal basis functions and employing M-estimation to mitigate outlier effects, our approach improves the accuracy and reliability of parameter estimates in the spatial functional linear regression context. Simulation studies and empirical data analysis substantiate the effectiveness of our methods, while an appendix explores the Fisher consistency and influence function of the FPCA-based approach. The \texttt{rfsac} package in \Rlogo{}\footnote{The \Rlogo\ package \texttt{rfsac} is available at \url{https://github.com/UfukBeyaztas/rfsac}} implements these robust estimation strategies, ensuring practical applicability for researchers and practitioners.
\end{abstract}

\noindent \textit{Keywords}: functional principal component analysis; functional partial least squares; M-estimation; spatial autoregressive model; spatial dependence. 

\newpage
\spacingset{1.4}

\section{Introduction} \label{sec:1}

Since the seminal paper by \cite{Hastie1993}, which delves into a functional linear regression model with a scalar response and functional predictor (SoFRM), a multifaceted landscape of functional linear models has emerged in the statistical domain. These models have been intricately developed to accommodate an assortment of statistical intricacies, catering to diverse research intentions and methodological nuances \citep[see, e.g.,][]{Preda2005, Ramsay2006, Morris2015, Reiss2017}.

Consider a random sample $\lbrace Y_i, \X_i(t): i = 1, \ldots, n \rbrace$ drawn from the joint distribution $(Y, \X)$. Here, $Y \in \mathbb{R}$ assumes the role of a continuous scalar random variable, while $\X = \left\lbrace \X(t) \right\rbrace_{t \in \mathcal{I}}$ constitutes a stochastic process characterized by intricate temporal dynamics. The components of this process are portrayed as trajectories within the $\mathcal{L}^2(\mathcal{I})$ Hilbert space, denoted by $\mathcal{H}$, where $\mathcal{I}$ denotes a closed interval that rigorously delineates the spatial domain in which these trajectories exist. In this context, the scalar-on-function regression model (SoFRM) takes the form:
\begin{equation}\label{eq:sofrm}
Y = \beta_0 + \int_{\mathcal{I}} \X(t) \beta(t) dt + \epsilon,
\end{equation}
where $\beta_0 \in \mathbb{R}$ is the intercept, $\beta(t) \in \mathcal{L}^2(\mathcal{I})$ denotes the regression coefficient function within the square-integrable space, and $\epsilon$ stands for the random error term, typically assumed to be independent and identically distributed (i.i.d.) as a Gaussian real-valued random variable with zero mean and variance $\sigma^2$.

The linear SoFRM in~\eqref{eq:sofrm} has been extended into various model types, including nonlinear SoFRM \citep[see, e.g.,][]{James2005, Yao2010, Muller2013} and nonparametric SoFRM \citep[see, e.g.,][]{Reiss2017}. A common assumption across these models is the element-wise independence within the sampled data. However, in several domains such as demography, economics, environmental sciences, agronomy, and mining, spatially correlated data sets present significant challenges.

A number of methodologies have been developed for analyzing data derived from discrete data matrices with spatial dependence \citep[see, e.g.][]{Anselin1998, Lesage2009, Schabenberger2017}. Furthermore, significant research effort has been devoted to the analysis of spatially correlated functional data \citep[see, e.g.,][]{Nerini2010, Caballero2013, Zhang2016, Menafoglio2017, Giraldo2018, Kuenzer2021}. However, the methodologies developed in these studies predominantly utilize kriging methods tailored for point-referenced data. This study specifically focuses on the application of the SoFRM to basic areal unit data.

In spatial linear regression models for areal data, three primary frameworks are commonly employed to analyze spatial dependence: the spatial autoregressive model, the spatial error model, and the spatial Durbin model \citep{Lesage2009}. Among these, the spatial autoregressive model serves as a cornerstone, with advancements in this model readily applicable to the other two \citep{Huang2021}. Compared to the spatial error and spatial Durbin models, the spatial autoregressive model offers a more straightforward interpretation, as it relies on a single spatial dependence parameter that characterizes the spatial lag of the response to induce dependence among neighboring regions. Therefore, our study focuses on the spatial autoregressive scalar-on-function regression model (SSoFRM) to address spatial dependencies in functional areal data. Various estimation procedures for SSoFRMs have been explored by several researchers, including \cite{Medina2011}, \cite{Medina2012}, \cite{Pineda2019},  \cite{Huang2021}, \cite{Hu2021}, and \cite{Bouka2023}.

The estimation methods used in the aforementioned studies predominantly rely on maximum likelihood-type estimators, which are known for their sensitivity to outliers. In practical applications, outliers are frequently encountered and can substantially affect these estimators, resulting in biased parameter estimates and less reliable predictive outcomes. The vulnerability of maximum likelihood estimators to outliers in spatial data has been illustrated by \cite{Luna2002}. To date, there remains a gap in developing a robust approach for accurately estimating the spatial lag parameter and the regression coefficient function within the framework of SSoFRM.

This study bridges the gap by introducing two robust methods for estimating the SSoFRM in the presence of outliers, discussing the relative advantages and disadvantages of each approach. The distinction between the methods lies in their treatment of the infinite-dimensional functional predictor projected onto a space defined by orthonormal basis functions. The first method utilizes the robust functional principal component (RFPC) approach by \cite{Bali2011}, while the second draws on the robust functional partial least squares (RFPLS) approach outlined by \cite{Beyaztas2022}, rooted in the partially robust M-regression method of \cite{Serneels2005}. By decomposing the functional predictor using RFPC and RFPLS, we approximate the infinite-dimensional SSoFRM within a finite-dimensional space spanned by basis expansion coefficients. This transformation allows us to convert the SSoFRM into a finite-dimensional spatial autoregressive model based on these coefficients for the scalar response. We then apply the M-estimator framework proposed by \cite{Tho2023} to estimate the parameters of this model efficiently.

Within a regression context, two types of outliers can arise: 
\begin{inparaenum}
\item[1)] leverage points, which denote uncommon observations within the predictor space, and 
\item[2)] vertical outliers, indicating unusual observations within the response variable. 
\end{inparaenum}
Both proposed methods incorporate mechanisms to mitigate the impact of leverage points using the RFPC and RFPLS techniques. Additionally, the M-estimator is utilized to reduce the influence of vertical outliers during parameter estimation within the finite-dimensional space. Therefore, both methods demonstrate robustness against leverage points in predictor variables and vertical outliers in the response variable.

The methodology based on RFPC draws on insights from the studies by \cite{kalogridis2019} and \cite{Tho2023}, enabling the establishment of Fisher-consistency and asymptotic consistency for the proposed estimators under specified regulatory conditions. The influence functions of these estimators are derived by combining the influence functions of RFPC with those of the M-estimator introduced by \cite{Tho2023}.

In their seminal work, \cite{Delaigle2012} introduces a compelling alternative formulation for the FPLS mechanism, articulated entirely in terms of functions. This innovative approach establishes an equivalent space that mirrors the dimensionality inherent in traditional partial least squares methods. Their study not only confirms the consistency of estimation and precision but also broadens the scope of functional partial least squares. In contrast, the RFPLS approach requires the application of case weights that undergo iterative updates to mitigate the influences of leverage points and vertical outliers. This complexity prevents RFPLS from being represented solely in functional terms, necessitating an empirically driven iterative approximation akin to standard partial least squares. Consequently, the implicit nature of the RFPLS algorithm presents challenges in deriving explicit theoretical outcomes \citep[see][for further insights]{Delaigle2012}.

The structure of this paper unfolds as follows. Section~\ref{sec:2} introduces the regression model and defines its parameters. Section~\ref{sec:3} details the methodology for parameter estimation. Given the challenge of infinite dimensionality, the function-valued regression coefficient function is estimated through basis function expansions, outlined in Sections~\ref{sec:3.1} and~\ref{sec:3.2}. Section~\ref{sec:3.3} discusses the application of the M-estimate for robust parameter estimation. Empirical data analysis using our approach is discussed in Section~\ref{sec:4}, where the results are detailed. Section~\ref{sec:5} concludes the paper, offering insights into potential extensions of the methodology. Additionally, in the Appendix, we provide a detailed demonstration of the Fisher consistency of the newly proposed RFPC-based estimator. Furthermore, the influence function of this estimator is derived by leveraging the influence functions of both RFPC and the M-estimator as described by \cite{Tho2023}. This analysis underscores that the proposed estimator is predominantly influenced by robust leverage points. In an online supplement file, a series of simulation studies is presented to validate the proposed method.

\section{Model and notations}\label{sec:2}

We begin by introducing some notations that will be utilized throughout this study. Consider the Hilbert space $\mathcal{H}$, defined on the probability space $(\Omega, \mathcal{A}, \mathcal{F})$, where the norm $\Vert \alpha \Vert^2$  is the square of the inner product $\langle \alpha, \alpha \rangle$. Let $\mathcal{I} \subseteq \mathbb{R}$ be a closed and bounded interval, and $\mathcal{L}^2(\mathcal{I})$ denote the space of square-Lebesgue-integrable functions on $\mathcal{I}$, meaning that for any $f \in \mathcal{L}^2(\mathcal{I})$, it holds that $\int_{\mathcal{I}} f^2(t) dt < \infty$. The inner product $\langle f, g \rangle$ is defined for all $f, g  \in \mathcal{L}^2(\mathcal{I})$ as $\langle f, g\rangle = \int_{\mathcal{I}} f(t) g(t) dt$. This setup ensures that $\mathcal{H}$ is a Hilbert space, providing a complete normed structure for the analysis of functions defined over the interval $\mathcal{I}$.

Let us turn our focus to a stochastic process described as $\lbrace Y_s, \X_s(t): s \in \mathcal{D} \subset \mathbb{R}^q, q \geq 1 \rbrace$, where $Y = (Y_1, \ldots, Y_n)^\top \in \mathbb{R}^n$ represents the scalar responses observed from discrete and unevenly spaced lattice subsets denoted by $\mathcal{D}$, consisting of spatial units $s_1, \ldots, s_n$, and $\X(t) = [\X_1(t), \ldots, \X_n(t)]^\top \in \mathcal{L}^2(\mathcal{I})$ is the corresponding vector of functional predictors. To simplify, we use $i$ to denote the spatial unit $s_i$. Accordingly, the SSoFRM takes the form:
\begin{equation}\label{eq:ssofrm}
Y = \beta_0 \bm{1}_n + \rho \bm{W} Y + \int_{\mathcal{I}} \X(t) \beta(t) dt + \epsilon,
\end{equation}
where $\beta_0 \in \mathbb{R}$ represents the intercept parameter, $\bm{1}_n$ denotes an $n$-dimensional vector of ones, $\rho \in (-1, 1)$ stands for the unknown spatial autocorrelation parameter, $\bm{W} = (w_{i i^{\prime}})_{1 \leq i, i^{\prime} \leq n}$ is an $n \times n$ spatial weight matrix, with elements $w_{i i^{\prime}}$ denoting the spatial relationship between locations $i$ and $i^{\prime}$, $\beta(t) \in \mathcal{L}^2(\mathcal{I})$ signifies the regression coefficient function, and $\epsilon \sim N(0, \Sigma)$, where $\Sigma$ denoted the variance-covariance matrix, represents the random error term. Note that the random variables in both~\eqref{eq:sofrm} and~\eqref{eq:ssofrm} represent similar entities, however, in SSoFRM the random variables are assumed to be observed from discrete and unevenly spaced lattice subsets $\mathcal{D} = \{s_1, \ldots, s_n \}$, with the spatial dependencies captured by the spatial weight matrix $\bm{W}$ and the spatial autocorrelation parameter $\rho$.

To illustrate the application of the SSoFRM, consider a simple case where $n = 3$ spatial units. Let the spatial domain $\mathcal{D}$ consist of three locations, indexed as $i_1$, $i_2$, and $i_3$. We choose $\mathcal{I} = [0,1]$ as the time interval over which the functional predictor $\X_i(t)$ is observed, where $t \in \mathcal{I}$. Let the spatial weight matrix $\bm{W}$ for these three locations be defined using an inverse distance weighting scheme. For simplicity, assume the following matrix: 
\[
\bm{W} = \begin{bmatrix}
0 & 0.4 & 0.4 \\
0.4 & 0 & 0.6 \\
0.4 & 0.6 & 0
\end{bmatrix},
\]
where each off-diagonal element represents the weight between two neighboring spatial units, normalized to sum to 1 in each row. The diagonal entries are set to $0,$ reflecting that a location does not influence itself. Suppose $\rho = 0.4$ and $\X_1(t) = \sin(2 \pi t)$, $\X_2(t) = \cos(2 \pi t)$, and $\X_3(t) = t(1-t)$. Let $Y_1$, $Y_2$, and $Y_3$ denote the corresponding scalar response at each spatial location $i$. Then, by applying the SSoFRM in~\eqref{eq:ssofrm}, the spatial dependence between the responses is captured by $\rho \bm{W}$, and the functional predictor $\X_i(t)$ is used to explain the variation in each response through the integral term.

In Model~\eqref{eq:ssofrm}, the spatial autocorrelation parameter $\rho$ gauges the strength of spatial dependence among neighboring locations. Larger values of $\rho$ indicate a substantial influence of neighboring observations on $Y_i$. When $\rho = 0$, the SSoFRM reduces to the SoFRM. The SSoFRM~\eqref{eq:ssofrm} is specifically characterized by the spatial weight matrix $\bm{W}$. The arrangement of the $n$ spatial units can be either regularly or irregularly distributed across any spatial region, provided the construction of the $\bm{W}$ matrix is feasible. In the context of data observed from a regular grid, units are considered neighbors if they share borders, corners, or both \citep{Anselin1998}. Conversely, in situations where dataset is observed from an irregular grid, units are deemed neighbors if they share common edges \citep{Huang2021}. The elements of a symmetric weight matrix $\bm{W}$, denoted by $w_{i i^{\prime}}$, can be constructed based on a distance metric, such that $w_{i i^{\prime}} = m(d_{i i^{\prime}})$, where $m(\cdot)$ is a monotonically decreasing function, and $d_{i i^{\prime}}$ represents the distance between locations $i$ and $i^{\prime}$. This distance could pertain to geographical, economic, social, policy-related factors, or a combination of these \citep[see, e.g.,][]{Yu2016}. Notably, the spatial weight matrix need not be symmetric; asymmetrical weight matrices can also find application in the model depending on the data characteristics, as demonstrated in \cite{Huang2021}. $\bm{W}$ is row-normalized, i.e., $w_{i i^{\prime}} = \frac{ m(d_{i i^{\prime}})}{\sum_{j^{\prime}=1}^n m(d_{i j^{\prime}})}$, to ensure that the sum of its elements within each row converges to unity, accompanied by enforcing zero values along the diagonal entries.

To gain deeper insight into the generation of the scalar response within the SSoFRM, Model~\eqref{eq:ssofrm} can be reformulated as follows:
\begin{equation}\label{eq:rexp}
Y = (\mathbb{I}_n - \rho \bm{W})^{-1} \beta_0 \bm{1}_n + (\mathbb{I}_n - \rho \bm{W})^{-1} \int_{\mathcal{I}} \X(t) \beta(t) dt + (\mathbb{I}_n - \rho \bm{W})^{-1} \epsilon,
\end{equation}
where $\mathbb{I}_n$ is an $n \times n$-dimensional diagonal matrix. Equation~\eqref{eq:rexp} unveils two important insights about the scalar response. Firstly, because the conditional expectation of the response, given the functional predictor, is not equal to $\int_{\mathcal{I}} \X(t) \beta(t) dt$, the spatial units' locations influence the mean of $Y$ through the functional predictor. Secondly, as indicated by the error term $(\mathbb{I}_n - \rho \bm{W})^{-1} \epsilon$, the residuals of $Y$ exhibit spatial correlation \citep{Huang2021}.

\section{Estimation}\label{sec:3}

The proposed methodology comprises two sequential steps. Initially, the infinite-dimensional functional predictor $\X(t)$ undergoes projection onto a finite-dimensional space spanned by a set of orthonormal basis of functions. Moreover, the regression coefficient function $\beta(t)$ is also represented within these orthonormal basis functions, enabling the representation of the integral component in~\eqref{eq:ssofrm}, specifically $\int_{\mathcal{I}} \X(t) \beta(t) dt$, through finite real vectors. This transformation converts the infinite-dimensional SSoFRM into a finite-dimensional spatial autoregressive model of scalar response $Y$ on the basis coefficients. Subsequently, we apply the M-estimator proposed by \cite{Tho2023} to estimate the regression parameters, i.e., $\beta_0$, $\rho$, and $\beta(t)$.

Let us now suppose that both $\X(t)$ and $\beta(t)$ are generated by the same orthonormal basis functions $\lbrace \phi_1(t), \phi_2(t), \ldots \rbrace \in \mathcal{L}^2(\mathcal{I})$. Then, $\X(t)$ and $\beta(t)$ can be expressed in terms of the basis expansion as follows:
\begin{equation*}
\X(t) = \sum_{k \geq 1} a_k \phi_k(t), \qquad \beta(t) = \sum_{k \geq 1} \beta_k \phi_k(t),
\end{equation*}
where $a_k = \int_{\mathcal{I}} \X(t) \phi_k(t) dt$ and $\beta_k = \int_{\mathcal{I}} \beta(t) \phi_k(t) dt$. Subsequently, leveraging the orthonormality condition (i.e., $\int_{\mathcal{I}} \bm{\phi}(t) \bm{\phi}^\top(t) dt = 1$, where $\bm{\phi}(t) = [\phi_1(t), \phi_2(t), \ldots]^\top$), the SSoFRM in~\eqref{eq:ssofrm} can be reformulated as follows:
\begin{equation}\label{eq:fdr}
Y = \beta_0 \bm{1}_n + \rho \bm{W} Y + \sum_{k \geq 1} a_k \beta_k + \epsilon.
\end{equation}
In practical applications, the variability in data is often represented using a finite number of orthonormal basis functions. Therefore, we proceed by expanding the functional predictor and regression coefficient function using a predefined truncation constant $K$. Thus, for the model in~\eqref{eq:fdr}, the approximation is formulated as:
\begin{equation}\label{eq:tfdr}
Y \approx \beta_0 \bm{1}_n + \rho \bm{W} Y + \sum_{k = 1}^K a_k \beta_k + \epsilon.
\end{equation}

Two notable examples of orthonormal bases are the functional principal component (FPC) basis \citep{Dauxois1982} and the FPLS basis \citep{Preda2005}. For detailed insights into FPC and FPLS estimates within the SSoFRM framework, we refer to \cite{Huang2021}. In the FPC approach, orthonormal bases are computed by maximizing the covariance between functional predictors, aiming to capture the principal modes of variation. Conversely, FPLS derives orthonormal bases by maximizing the covariance between the scalar response and functional predictors, emphasizing predictive efficiency. Both methods optimize the conventional covariance operator using a least-squares loss function, which is vulnerable to outliers. In the presence of outliers, the orthonormal bases derived from FPC and FPLS may inadequately capture the underlying data structure, leading to compromised basis expansion coefficients. To address this issue, we introduce two sets of robust orthonormal basis functions: RFPC and RFPLS. Throughout subsequent sections, RFPC quantities will be denoted with the superscript $^{(1)}$, while RFPLS quantities will be denoted with the superscript $^{(2)}$.

\subsection{RFPC basis}\label{sec:3.1}

The RFPC bases are obtained similarly to the classical FPC bases, but they utilize the M-scale functional, denoted as $\sigma_M$, instead of the traditional variance. For a random variable $X$ and a location parameter $\mu$, the M-scale functional is obtained by solving a continuous loss function $\varphi_1: \mathbb{R} \rightarrow \mathbb{R}$, such that $\mathbb{E}\lbrace \varphi_1[ (X - \mu) / \sigma_M ] \rbrace = \delta$. The M-estimate of scale for a given sample $\bm{X} = \lbrace X_1, \ldots, X_n \rbrace$, denoted as $\widehat{\sigma}_M$, is derived by solving the M-estimating equation:
\begin{equation*}
\frac{1}{n} \sum_{i = 1}^n \varphi_1 \left(\frac{X_i - \widehat{\mu}}{\widehat{\sigma}_M} \right) = \delta,
\end{equation*}
where $\widehat{\mu}$ represents the location estimate of $\bm{X}$. In this study, following the approach similar to \cite{Bali2011}, we utilize the loss function introduced by \cite{Beaton1974} to compute the M-estimate of scale. It is expressed as:
\begin{equation}\label{eq:loss1}
        \varphi_{1,c}(u) =
        \left\{ \begin{array}{ll}
            \frac{u^2}{2} \left( 1 - \frac{u^2}{c^2} + \frac{u^4}{3c^4} \right)  & \text{if}~ \vert u \vert \leq c, \\
            \frac{c^2}{6} & \text{if}~ \vert u \vert > c,
        \end{array} \right.
\end{equation}
where $c$ serves as the tuning parameter that governs the robustness and efficiency of the estimator. Typically, commonly values chosen for $c$ and $\delta$ are $c = 1.56$ and $\delta = 0.5$, as utilized in our numerical analyses. These values ensure a 50\% breakdown point with respect to the normal distribution, as discussed in \cite{Bali2011}.

Let $\mathcal{F}$ and $\mathcal{F}[\alpha]$ denote the distributions of $\X$ and $\langle \alpha, \X \rangle$, respectively. The orthonormal RFPC bases corresponding to a pre-spepcified M-scale functional $\sigma_M(\mathcal{F})$ are defined as follows:
\[ \begin{cases}
\phi_k^{(1)}(\mathcal{F}) = \underset{\begin{subarray}{c}
      \Vert \alpha \Vert = 1
\end{subarray}}{\argmax}~ \sigma_M(\mathcal{F}[\alpha]), & k = 1, \\
\phi_k^{(1)}(\mathcal{F}) = \underset{\begin{subarray}{c}
      \Vert \alpha \Vert = 1, \alpha \in \mathcal{B}_k
\end{subarray}}{\argmax}~ \sigma_M(\mathcal{F}[\alpha]), & k \geq 2,
   \end{cases}
\]
where $\mathcal{B}_k = \left\lbrace \alpha \in \mathcal{L}^2(\mathcal{I}): \langle \alpha, \phi_{\ell}^{(1)} ( \mathcal{F} ) \rangle = 0, ~ 1 \leq \ell \leq \ (k - 1) \right\rbrace$, and the $k$\textsuperscript{th} largest eigenvalue, denoted by $\lambda_k$, is defined as follows:
\begin{equation*}
\lambda_k(\mathcal{F}) = \sigma_M^2(\mathcal{F}[\phi_k^{(1)}]) = \underset{\begin{subarray}{c}
      \Vert \alpha \Vert = 1, \ \alpha \in \mathcal{B}_k
\end{subarray}}{\max} \sigma_M^2(\mathcal{F}[\alpha]).
\end{equation*}

Let $\sigma_M$ be defined as $\sigma_M(\mathcal{F}_n[\alpha])$. Define $s^2_M: \mathcal{L}^2[\mathcal{I}] \rightarrow \mathbb{R}$ as the empirical M-scale functional, where $s^2(\alpha) = \sigma_M^2(\mathcal{F}[\alpha])$. In this context, the RFPC estimates of the orthonormal bases are derived as follows:
\[ \begin{cases}
\widehat{\phi}_k^{(1)}(t) = \underset{\begin{subarray}{c}
      \Vert \alpha \Vert = 1
\end{subarray}}{\argmax}~ s_M(\alpha), & k = 1, \\
\widehat{\phi}_k^{(1)}(t) = \underset{\begin{subarray}{c}
      \alpha \in \widehat{\mathcal{B}}_k
\end{subarray}}{\argmax}~ s_M(\alpha), & k \geq 2,
   \end{cases}
\]
where $\widehat{\mathcal{B}}_k = \left\lbrace \alpha \in \mathcal{L}^2 (\mathcal{I}): \Vert \alpha \Vert = 1, \langle \alpha, \widehat{\phi}_{\ell}^{(1)} \rangle = 0, ~ \forall~ 1 \leq \ell \leq \ (k - 1) \right\rbrace$. The eigenvalues are computed as follows:
\begin{equation*}
\widehat{\lambda}_k = s^2_M (\widehat{\phi}_k^{(1)}), \quad k \geq 1.
\end{equation*}
Once the orthonormal RFPC bases are obtained, the estimates of the basis expansion coefficients for $\X(t)$ and $\beta(t)$ are respectively given by $\widehat{a}_k^{(1)} = \int_{\mathcal{I}} \X(t) \widehat{\phi}_k^{(1)}(t) dt$ and $\widehat{\beta}_k^{(1)} = \int_{\mathcal{I}} \beta(t) \widehat{\phi}_k^{(1)}(t) dt$, for $k = 1, \ldots, K$. Subsequently, the sample counterpart of~\eqref{eq:tfdr} becomes:
\begin{equation}\label{eq:rfpc_c}
Y \approx \beta_0 \bm{1}_n + \rho \bm{W} Y + \sum_{k = 1}^K \widehat{a}_k^{(1)} \widehat{\beta}_k^{(1)} + \epsilon.
\end{equation}

\subsection{RFPLS basis}\label{sec:3.2}

The RFPLS method adopts an iterative procedure introduced by \cite{Preda2005} to compute basis functions and their orthogonal components. Unlike FPLS, RFPLS utilizes the partially robust M-regression algorithm developed by \cite{Serneels2005} to minimize the influence of outliers in both scalar responses and functional predictors during the derivation of basis functions and their corresponding components. Importantly, the computation of RFPLS bases excludes the spatial lag term $\rho \bm{W} Y$, ensuring that the basis functions are derived independently of spatial correlation considerations. The same approach for the classical FPLS method for the SSoFRM is suggested by \cite{Huang2021}.

Let $h = 1, 2, \ldots$ represent the iteration number. The algorithm for obtaining RFPLS basis functions and their corresponding orthogonal components is outlined in Algorithm~\ref{alg:fpls}.
\begin{algorithm}[!htb]
For $h = 1$, set $Y_{(h-1)} = Y$ and $\X_{(h-1)}(t) = \X(t)$. For $h = 1, \ldots, K$, repeat the following steps: \\
Compute orthonormal RFPLS basis functions $\phi^{(2)}_{(h)}(t)$ and define orthogonal components $a_{(h)}$ by solving Tucker's criterion:
\begin{equation*}
\phi^{(2)}_{(h)}(t) = \underset{\begin{subarray}{c}
  \phi^{(2)}(t) \in \mathcal{L}^2(\mathcal{I}) \\ \Vert \phi^{(2)}(t) \Vert = 1
  \end{subarray}}{\argmax} \text{Cov}_r^2 \big( Y_{(h-1)}, \int_{\mathcal{I}} \X_{(h-1)}(t) \phi^{(2)}(t) dt \big),
\end{equation*}
where $\text{Cov}_r(\cdot, \cdot)$ is a robust covariance operator. \\
Calculate orthogonal components as linear functionals of $\X_{(h-1)}(t)$ and $\phi^{(2)}_{(h)}(t)$
\begin{equation*}
a_{(h)} = \int_{\mathcal{I}} \X_{(h-1)}(t) \phi^{(2)}_{(h)}(t) dt.
\end{equation*} \\
Construct two individual regression models for $\X_{(h-1)}(t)$ and $Y_{(h-1)}$ as follows:
\begin{equation*}
Y_{(h-1)} =  q_{(h)} a_{(h)} + \epsilon_{(h)}^{y}, \qquad
\X_{(h-1)}(t) =  p_{(h)}(t) a_{(h)} + \epsilon_{(h)}^{x}(t),
\end{equation*}
where $q_{(h)} = \mathbb{E}[Y_{(h-1)} a_{(h)}]/\mathbb{E}[a_{(h)}^2]$ and $p_{(h)}(t) = \mathbb{E}[\X_{(h-1)}(t) a_{(h)}]/\mathbb{E}[a_{(h)}^2]$. \\
Update $Y_{(h)}$ and $\X_{(h)}(t)$ as $Y_{(h)} = \epsilon_{(h)}^{y}$ and $\X_{(h)}(t) = \epsilon_{(h)}^{x}(t)$. Terminate if $h = K$; otherwise, return to Step~2.
\caption{\small{FPLS components}}
\label{alg:fpls}
\end{algorithm}

Direct computation of RFPLS basis functions and their corresponding components is hindered by the infinite-dimensional nature of functional predictors. For practical purposes, given a sample $\lbrace Y_i, \X_i(t): i = 1, \ldots, n \rbrace$, the functional predictor is approximated within a finite-dimensional space using an appropriate basis expansion function such as B-spline or Fourier basis. This approximation is represented as $\X(t) = \sum_{m =1}^M \widetilde{a}_m \psi_m(t) = \widetilde{\bm{a}} \bm{\psi}^\top(t)$, where $M$ denotes the number of basis expansion functions. The value of $M$ can be determined using cross-validation, explained variance criterion, or information criteria such as the Bayesian Information Criterion. Additionally, from the spectral analysis of the covariance function in RFPLS, the RFPLS basis functions $\phi^{(2)}(t)$ and regression coefficient function $\beta(t)$ are expressed in the same basis, i.e., $\phi^{(2)}(t) = \widetilde{\bm{\phi}} \bm{\psi}^\top(t)$ and $\beta(t) = \widetilde{\bm{\beta}} \bm{\psi}^\top(t)$. Consequently, the challenge of infinite-dimensional eigen-analysis in RFPLS regression of $Y$ on $\X(t)$ is transformed into a finite-dimensional eigen-analysis problem. By employing $\bm{\Psi} = \int_{\mathcal{I}} \psi(t) \psi^\top(t) dt$ to denote the matrix of inner products between basis functions and $\bm{\Psi}^{1/2}$ as the square root of $\Psi$, it was demonstrated by \cite{Beyaztas2022} that the FPLS regression of $Y$ on $\X(t)$ equates to the finite-dimensional PLS regression of $Y$ on $\widetilde{\bm{a}} (\bm{\Psi}^{1/2})^\top$, ensuring that both models yield identical PLS components at each step of the algorithm.

To solve the regression problem of $Y$ on $\bm{D} = \widetilde{\bm{a}} (\bm{\Psi}^{1/2})^\top$, the RFPLS method incorporates robust eigenvectors denoted as $\widetilde{\phi}$, which utilizes a weighted covariance function to optimize the objective function:
\begin{equation*}
\widetilde{\phi} = \text{Cov}_{r}(Y, \bm{D})  \Leftrightarrow \underset{\begin{subarray}{c}
  \widetilde{\phi}	
  \end{subarray}}{\argmin} \sum_{i=1}^n r_i^2 (Y_i - \bm{D}_i \widetilde{\phi} )^2,
\end{equation*}
where $r_i^2 = r_i^x r_i^y$ represents the case weight. $r_i^x = \omega_r(e_i / \widehat{\sigma})$ and $r_i^y = \omega_a(\Vert \bm{a}_i - \text{med}_k(\bm{a}_k) \Vert / \text{med}_i \Vert \bm{a}_i - \text{med}_k(\bm{a}_k) \Vert$ serve as symbolic representations of the weighting assigned to potential outliers in the predictor space and the response variable domain, respectively. Here, $\text{med}(\cdot)$ denotes the median operator, and $e_i$ represents the residual derived from the regression model involving scalar response with respect to the RFPLS expansion coefficients. It is essential to note that $\omega(x)$ denotes the Hampel weighting function, a key aspect elaborated in previous literature \citep[see, for example,][for a comprehensive exposition]{Serneels2005, Beyaztas2022}. 

In RFPLS, the case weights are dynamic rather than fixed, evolving through successive updates via an iterative algorithm akin to iterative reweighted partial least squares. Moreover, $\widehat{\widetilde{\bm{\phi}}}$ denotes the collection of eigenvectors obtained from the RFPLS technique. Utilizing $\widehat{\widetilde{\bm{\phi}}}$, the RFPLS basis functions are computed as $\widehat{\phi}^{(2)}(t) = \widehat{\widetilde{\bm{\phi}}} \bm{\psi}^\top(t)$. This leads us to the associated components encapsulated within $\widehat{\bm{a}}^{(2)} = \bm{D} \widetilde{\phi}$. Addressing the regression problem $Y = \gamma_0 + \widehat{\bm{a}}^{(2)} \bm{\gamma}$, where $\widehat{\bm{\gamma}}$ estimates $\gamma$, the basis expansion coefficients for the regression coefficient function are derived as $\widehat{\bm{\beta}}^{(2)} = (\bm{\Psi}^{-1/2})^\top \widehat{\widetilde{\bm{\phi}}} \widehat{\bm{\gamma}}$. Finally, the $K$-sets of quantities derived from the RFPLS algorithm, $\lbrace \widehat{a}^{(2)}_1, \ldots, \widehat{a}^{(2)}_K \rbrace$ and $\lbrace \widehat{\beta}^{(2)}_1, \ldots, \widehat{\beta}^{(2)}_K \rbrace$, are used to establish a truncated form similar to~\eqref{eq:rfpc_c} for the RFPLS method:
\begin{equation}\label{eq:rfpls_c}
Y \approx \beta_0 \bm{1}_n + \rho \bm{W} Y + \sum_{k = 1}^K \widehat{a}_k^{(2)} \widehat{\beta}_k^{(2)} + \epsilon.
\end{equation}

\subsection{M-estimate}\label{sec:3.3}

We now shift our focus to the approximate models~\eqref{eq:rfpc_c} or~\eqref{eq:rfpls_c}, specifically aiming to estimate the model parameters $\beta_0$, $\rho$, $\widehat{\beta}_k^{(1)}$ (or $\widehat{\beta}_k^{(2)}$), and the variance of the model, denoted by $\sigma^2$. For simplicity, we omit the superscripts $^{(1)}$ and $^{(2)}$ and concentrate solely on the parameter estimates.

Let us define the matrices $\bm{A} = (\widehat{a}_{i k})_{n \times K}$ and $\bm{Z} = [\bm{1}_n, \bm{A}]^\top$, where $\bm{1}_n$ is the vector of ones, and $\bm{\theta} = [\beta_0, \bm{\beta}^\top]^\top$, with $\bm{\beta} = [\widehat{\beta}_1, \ldots, \widehat{\beta}_K]$, representing the model parameters. Consequently, the matrix form of the approximate model~\eqref{eq:rfpc_c} (or~\eqref{eq:rfpls_c}) can be expressed as:
\begin{equation}\label{eq:appM}
Y \approx \rho \bm{W} Y + \bm{Z} \bm{\theta} + \epsilon.
\end{equation}
Defining the parameter vector for \eqref{eq:appM} as $\bm{\Theta} = [\bm{\theta}, \sigma, \rho]^\top$, the $\log$-likelihood function for $Y$ can be formulated as:
\begin{equation*}
\ell(\bm{\Theta}; Y) = -\frac{n}{2} \log(2 \pi) - n \log(\sigma) + \log \vert \det(\bm{I}_n - \rho \bm{W}) \vert - \frac{1}{2 \sigma^2} [ (\bm{I}_n - \rho \bm{W}) Y - \bm{Z} \bm{\theta} ]^\top [ (\bm{I}_n - \rho \bm{W}) Y - \bm{Z} \bm{\theta}].
\end{equation*}
The maximum likelihood estimator, denoted by $\widehat{\bm{\Theta}}_{ML}$, is obtained by maximizing this $\log$-likelihood function, i.e., $\widehat{\bm{\Theta}}_{ML} = \argmax_{\bm{\Theta}} \ell(\bm{\Theta}; Y)$. The estimating equation function of the maximum likelihood is represented as $\eta_{ML}(\bm{\Theta}; Y) = \partial\ell(\bm{\Theta}; Y)/\partial \bm{\Theta} = \bm{0}$, yielding the following estimating equations:
\begin{equation*}
\eta_{ML}(\bm{\Theta}; Y) = 
\begin{bmatrix}
    \frac{1}{\sigma^2} \bm{Z}^\top [ (\bm{I}_n - \rho \bm{W}) Y - \bm{Z} \bm{\theta}]\\
    \frac{1}{\sigma^3} [ (\bm{I}_n - \rho \bm{W}) Y - \bm{Z} \bm{\theta} ]^\top [ (\bm{I}_n - \rho \bm{W}) Y - \bm{Z} \bm{\theta}] - \frac{n}{\sigma} \\
    \frac{1}{\sigma^2} (\bm{W} Y)^\top [ (\bm{I}_n - \rho \bm{W}) Y - \bm{Z} \bm{\theta}] - \text{trace}[\bm{W}(\bm{I}_n - \rho \bm{W})^{-1}]
  \end{bmatrix}.
\end{equation*}

The influence of outliers in the predictor space on the maximum likelihood estimator is mitigated through the projection of the functional predictor onto a finite-dimensional space using robust decomposition methods (RFPC or RFPLS). However, as highlighted by \cite{Lee2004} and \cite{Tho2023}, the estimating function $\eta_{ML}(\bm{\Theta}; Y)$ lacks boundedness in $Y$, and consequently, in $\epsilon = (\bm{I}_n - \rho \bm{W}) Y - \bm{Z} \bm{\theta}$. This implies that the influence function of the maximum likelihood estimator remains unbounded in both $Y$ and $\epsilon$, making it vulnerable to outliers in both the response and error terms. To address this issue and obtain a robust estimator for $\bm{\Theta}$ that is resilient against outliers in the predictor space, response variable, and error term, we adopt the M-estimator proposed by \cite{Tho2023}.

Let us denote the M-estimator as $\widehat{\bm{\Theta}}_{R}$, which is obtained by solving $\eta_{R}(\bm{\Theta}; Y) = \bm{0}$. The corresponding estimating equations for the M-estimator are elaborated by \citep{Tho2023} as follows:
\begin{equation*}
\eta_R(\bm{\Theta}; Y) = 
\begin{bmatrix}
    \bm{Z}^\top \bm{\varphi}_{2, c_1}(\bm{\epsilon^*})\\
    \bm{\varphi}_{2, c_2}^\top(\bm{\epsilon^*}) \bm{\varphi}_{2, c_2}(\bm{\epsilon^*}) - n \widetilde{\varrho}(c_2) \\
    \frac{1}{\sigma} [\bm{G}(\rho) \bm{Z} \bm{\theta}]^\top \bm{\varphi}_{2, c_3}(\bm{\epsilon^*}) + \bm{\varphi}_{2, c_3}^\top(\bm{\epsilon^*}) \bm{G}^\top(\rho) \bm{\varphi}_{2, c_3}(\bm{\epsilon^*}) - \text{trace}[\bm{G}(\rho)] \widetilde{\varrho}(c_3)
  \end{bmatrix},
\end{equation*}
where $\bm{\epsilon^*} = [(\bm{I}_n - \rho \bm{W}) Y - \bm{Z} \bm{\theta}] / \sigma$ denotes the standardized residuals, $\bm{\varphi}_{2, c}: \mathbb{R} \rightarrow \mathbb{R}$ symbolizes the Huber function with tuning parameter $c > 0$ as $\varphi_{2,c}(\epsilon^*_i) = \epsilon^*_i \min \{1, c / \vert \epsilon^*_i \vert \}$, $\bm{G}(\rho) = \bm{W} (\bm{I}_n - \rho \bm{W})^{-1}$, $\widetilde{\varrho}(c) = 2 c^2 [1 - \Pi(c)] - 2 c \upsilon(c) - 1 + 2 \Pi(c)$ using the standard normal probability density function $\upsilon(\cdot)$ and cumulative distribution function $\Pi(\cdot)$, and where $c_1$, $c_2$, and $c_3$ serve as tuning parameters.

Compared to the estimating function $\eta_{ML}$ of maximum likelihood, the robust M-estimator applies Huber functions to the elements of the standardized residual vector, thereby ensuring boundedness of $\eta_{R}(\bm{\Theta}; Y)$ in both $Y$ and $\bm{\epsilon}$, consequently establishing boundedness in the influence function of the robust M-estimator. As discussed by \cite{Tho2023}, the robust M-estimator may not achieve the same efficiency as the maximum likelihood estimator, as a trade-off between robustness and efficiency exists, controlled by the tuning parameters $(c_1$, $c_2$, $c_3)$. As these tuning parameters approach infinity, the behaviour of the robust M-estimator converges with that of the maximum likelihood estimator. In our numerical analyses, we adopt the tuning parameter values $(c_1 = 1.4$, $c_2 = 2.4$, $c_3 = 1.65)$, yielding approximately 95\% efficiency as reported by \cite{Tho2023}. Obtaining the robust M-estimates requires an iterative algorithm, outlined in Algorithm~\ref{alg:estalg}.

\begin{algorithm}[!htb]
Initialize $\widehat{\bm{\Theta}}^{[0]} = [\widehat{\bm{\theta}}^{[0]}, \widehat{\sigma}^{[0]}, \rho^{[0]}]$ (e.g., maximum likelihood estimates), and the tuning parameters $(c_1, c_2, c_3) = (1.4, 2.4, 1.65)$. \\
At the $h$\textsuperscript{th} iteration, update the regression coefficient as follows:
\begin{equation*}
\widehat{\bm{\theta}}^{[h+1]} = [\bm{Z}^\top \bm{D}_w(\widehat{\bm{\Theta}}^{[h]}; Y) \bm{Z}]^\top \bm{Z}^\top \bm{D}_w(\widehat{\bm{\Theta}}^{[h]}; Y) (Y - \rho^{[h]} \bm{W} Y), 
\end{equation*}
where $\bm{D}_w(\widehat{\bm{\Theta}}^{[h]}; Y) = \text{diag}[\omega_1(\widehat{\bm{\Theta}}^{[h]}; Y), \ldots, \omega_n(\widehat{\bm{\Theta}}^{[h]}; Y)]$, $\omega_i(\widehat{\bm{\Theta}}^{[h]}; Y) = \varrho_{c_1}[\epsilon^*_i(\widehat{\bm{\Theta}}^{[h]}; Y)] \epsilon^*_i(\widehat{\bm{\Theta}}^{[h]}; Y)^{-1}$ if $\epsilon^*_i(\widehat{\bm{\Theta}}^{[h]}; Y) \neq 0$ and $\omega_i(\widehat{\bm{\Theta}}^{[h]}; Y) = 1$ if $\epsilon^*_i(\widehat{\bm{\Theta}}^{[h]}; Y) = 0$. \\
Update the scale estimate as follows:
\begin{equation*}
\widehat{\sigma}^{[h+1]} = \Big[ \frac{\widehat{\sigma}^{[h]}}{n \widetilde{\varrho}(c_2)} \bm{\varphi}_{2, c_2}[\bm{\epsilon^*}(\widehat{\bm{\theta}}^{[h]}, \widehat{\sigma}^{[h]}, \rho^{[h]}; Y)]^\top \bm{\varphi}_{2, c_2} [\bm{\epsilon^*}(\widehat{\bm{\theta}}^{[h]}, \widehat{\sigma}^{[h]}, \rho^{[h]}; Y)] \Big]^{1/2}.
\end{equation*} \\
Update the estimate of spatial autocorrelation by solving a one-dimensional optimization problem:
\begin{equation*}
\rho^{[h+1]} = \argmin_{\rho \in (u,v)} [\eta_R(\rho; \widehat{\bm{\theta}}^{[h+1]}, \widehat{\sigma}^{[h+1]}, Y)]^2.
\end{equation*} \\
Repeat Steps~2 to~4 until convergence, where convergence occurs when $\Vert \widehat{\bm{\Theta}}^{[h+1]} - \widehat{\bm{\Theta}}^{[h]} \Vert_2 < \varepsilon$ for a small $\varepsilon$.
\caption{\small{Iterative estimation algorithm for the robust M-estimator}}
\label{alg:estalg}
\end{algorithm}

Note that in the M-estimation algorithm, the invertibility of the matrix $(\bm{I}_n - \rho \bm{W})$  is critical, as it directly affects both the score function $\eta_{ML}(\bm{\Theta}; Y)$ and its robust version $\eta_R(\bm{\Theta}; Y)$. For $(\bm{I}_n - \rho \bm{W})$ to be invertible, the parameter $\rho$  must lie within a range that ensures this property. In Step 4 of the Algorithm~\ref{alg:estalg}, $\rho$ is constrained to lie within a range $(u, v) = (- \vert \lambda_{\text{min}}(\bm{W}) \vert^{-1}, 1)$, where $\lambda_{\text{min}}(\bm{W})$ denotes the smallest eigenvalue of the row-normalized $\bm{W}$. This constraint ensures that  $(\bm{I}_n - \rho \bm{W})$ remains invertible. However, even if $(\bm{I}_n - \rho \bm{W})$  is close to singular, it could negatively affect the algorithmic convergence, especially since $(\bm{I}_n - \rho \bm{W})^{-1}$ appears in both the score function and its robust version. To mitigate potential numerical issues related to near-singularity, a regularization approach could be considered. One option is to ensure that $\rho$ is estimated within a safe range by incorporating a constraint during the estimation process. Alternatively, adding a small regularization term, such as $\epsilon \bm{I}_n$, to the matrix $(\bm{I}_n - \rho \bm{W})$,  resulting in $(\bm{I}_n - \rho \bm{W} + \epsilon \bm{I}_n)$, could help stabilize the inversion, especially when $\rho$ approaches problematic values.

Let $\widehat{\bm{\Theta}} = [\widehat{\bm{\theta}}, \widehat{\sigma}, \widehat{\rho}]^\top$ denote the robust M-estimates, where $\widehat{\bm{\theta}} = [\widehat{\beta}_0, \widehat{\bm{\beta}}^\top]^\top$ represents the model parameters obtained using Algorithm~\ref{alg:estalg}. Consequently, the robust estimate of the regression coefficient function $\beta(t)$ is given by $\widehat{\beta}(t) = \sum_{k=1}^K \widehat{\beta}_k \widehat{\phi}_k(t)$. This leads to the robust estimation of the SSoFRM:
\begin{equation*}
\widehat{Y} = (\mathbb{I}_n - \widehat{\rho} \bm{W})^{-1} \widehat{\beta}_0 \bm{1}_n + (\mathbb{I}_n - \widehat{\rho} \bm{W})^{-1} \int_{\mathcal{I}} \X(t) \widehat{\beta}(t) dt.
\end{equation*}
The asymptotic consistency and influence function of $\widehat{\beta}$ obtained using RFPC decomposition are in the Appendix.

\section{Application to U.S. COVID-19 data}\label{sec:4}

\begin{figure}[!htb]
\centering
\includegraphics[width=8.7cm]{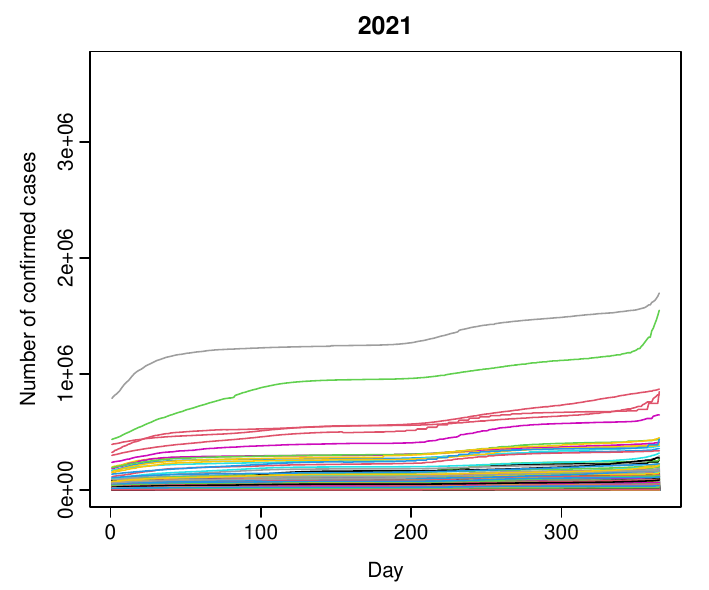}
\qquad
\includegraphics[width=8.7cm]{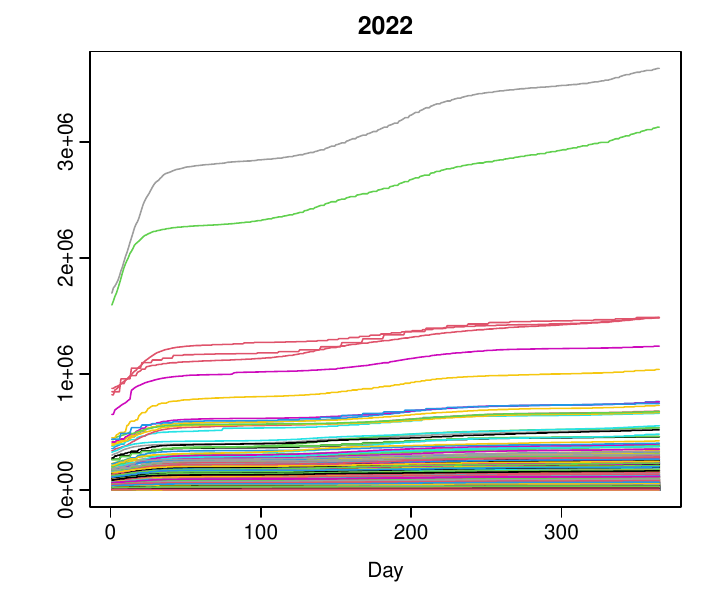}
\\ 
\includegraphics[width=8.7cm]{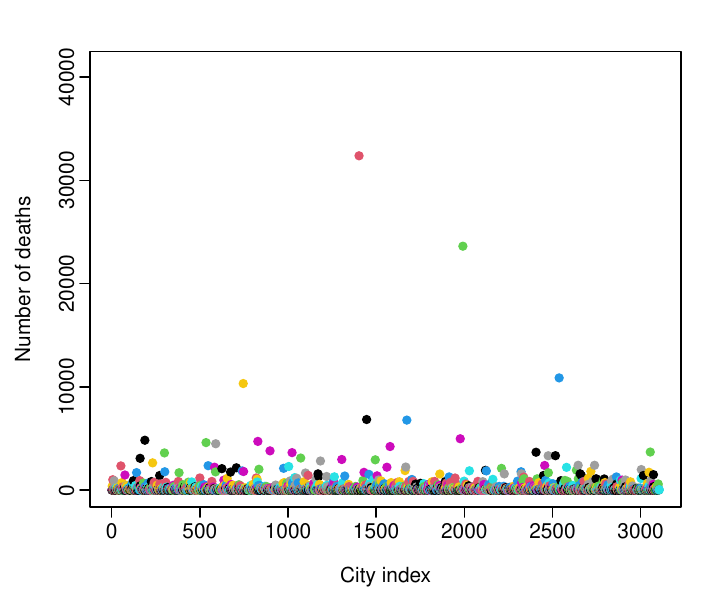}
\qquad
\includegraphics[width=8.7cm]{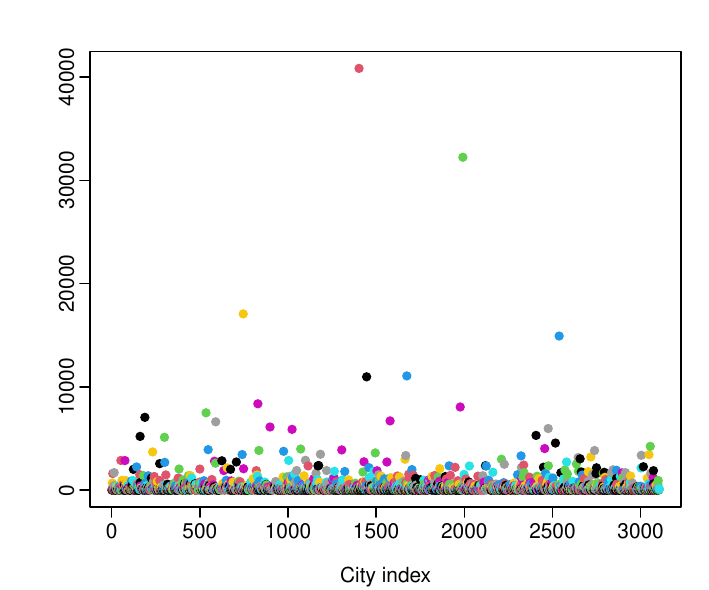}
\caption{\small{The graphical display of U.S. COVID-19 data presents the confirmed cases in the top panels and the average number of deaths in the bottom panels for 2021 (left panels) and 2022 (right panels). Different colors represent different cities, with the observations of confirmed cases being functions of days, i.e., $1 \leq t \leq 365$}.}\label{fig:Fig_1}
\end{figure}

The proposed RFPC and RFPLS-based robust estimation methods have been utilized within the SSoFRM framework to explore the relationship between COVID-19-related variables. In this study, $Y$ represents the average number of deaths (scalar response), while $\X(t)$ stands for confirmed cases (functional predictor) across the U.S. To facilitate this analysis, we collected data sets detailing the average number of deaths and confirmed COVID-19 cases for the years 2021 and 2022 from 3106 cities in the U.S. This data set was sourced utilizing the \texttt{COVID19} package \citep{COVID19}. In the preprocessing step, the data from the 3,106 cities was organized such that each city’s functional predictor (e.g., the time series of confirmed COVID-19 cases) was stacked as a separate row in the data matrix. For each city $i$, the scalar response (e.g., the average number of COVID-19-related deaths) and the corresponding functional predictors were aligned, allowing the model to account for spatial dependencies across cities through the spatial weight matrix $\bm{W}$. The stacking procedure ensured that both the functional data and scalar responses for all cities were considered simultaneously in the regression model. A graphical representation of the average number of deaths and confirmed COVID-19 cases for 2021 and 2022 across these cities in the U.S. is presented in Figure \ref{fig:Fig_1}.

Recently, several regression models have emerged, some of which have been applied to COVID-19 data sets to elucidate the relationship between COVID-19 infection-induced fatalities and various other indicators \citep[see, e.g.,][]{Giordano21, Acal21, BHA2023}. However, many of these models lack robustness in handling outliers. Although studies such as \cite{BHA2023} have introduced functional regression models to robustly investigate the dynamics of COVID-19 prevalence, to our knowledge, a reliable functional regression model that accounts for spatial dependence in the response variable elements to explore the dynamics of COVID-19 prevalence has yet to be established.

\begin{figure}[!htb]
\centering
\includegraphics[width=8.9cm]{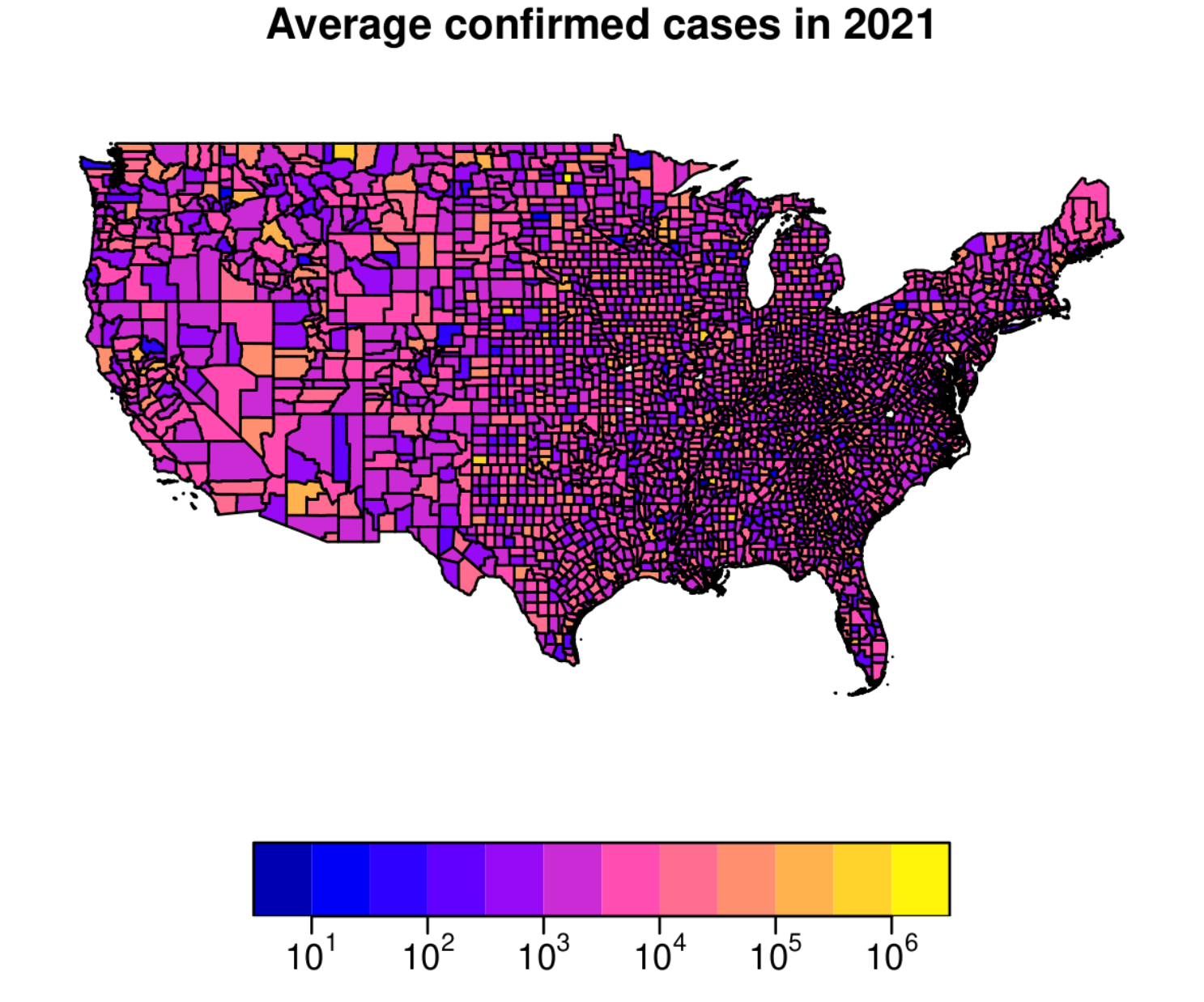}
\includegraphics[width=8.9cm]{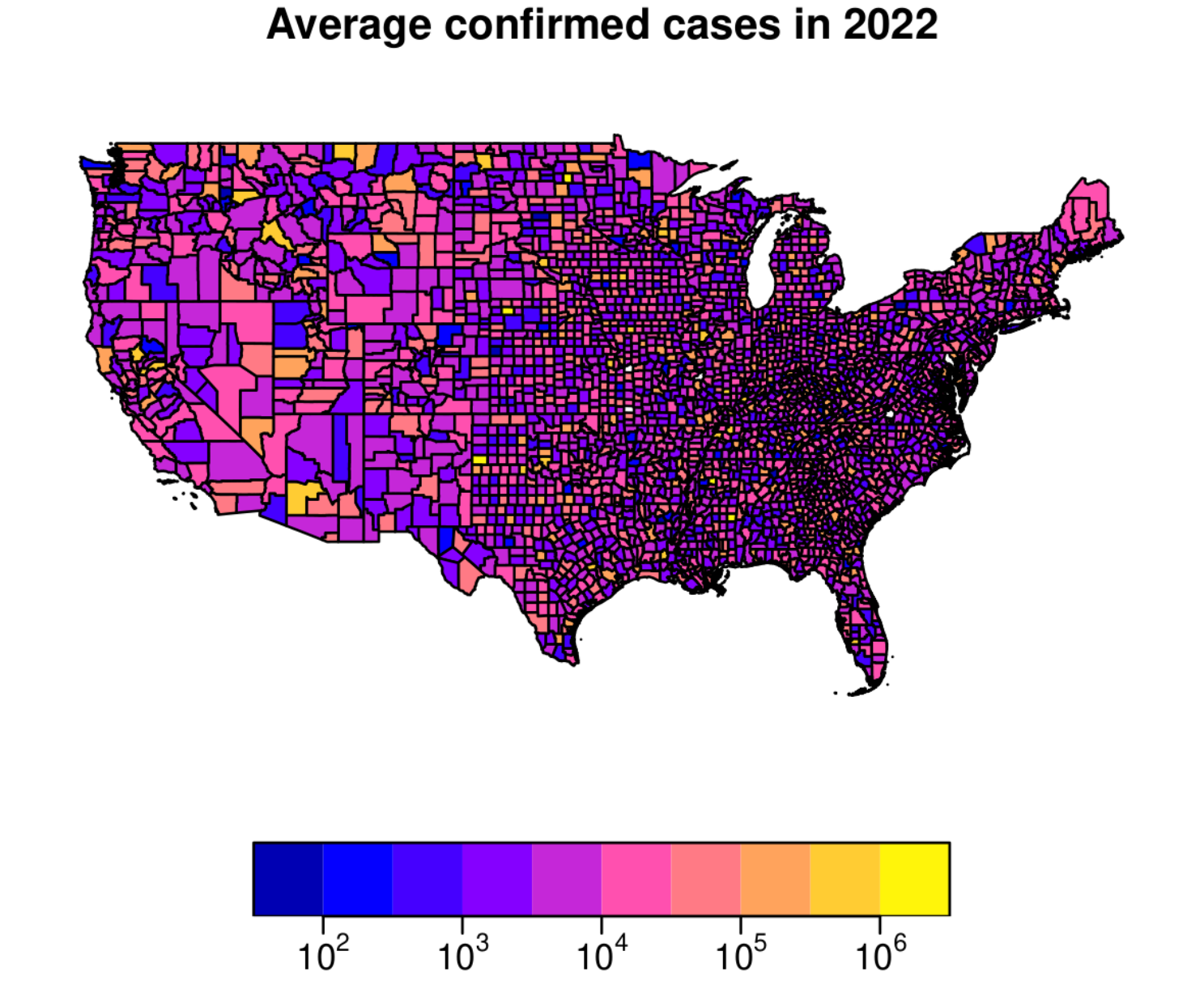}
\\ 
\includegraphics[width=8.9cm]{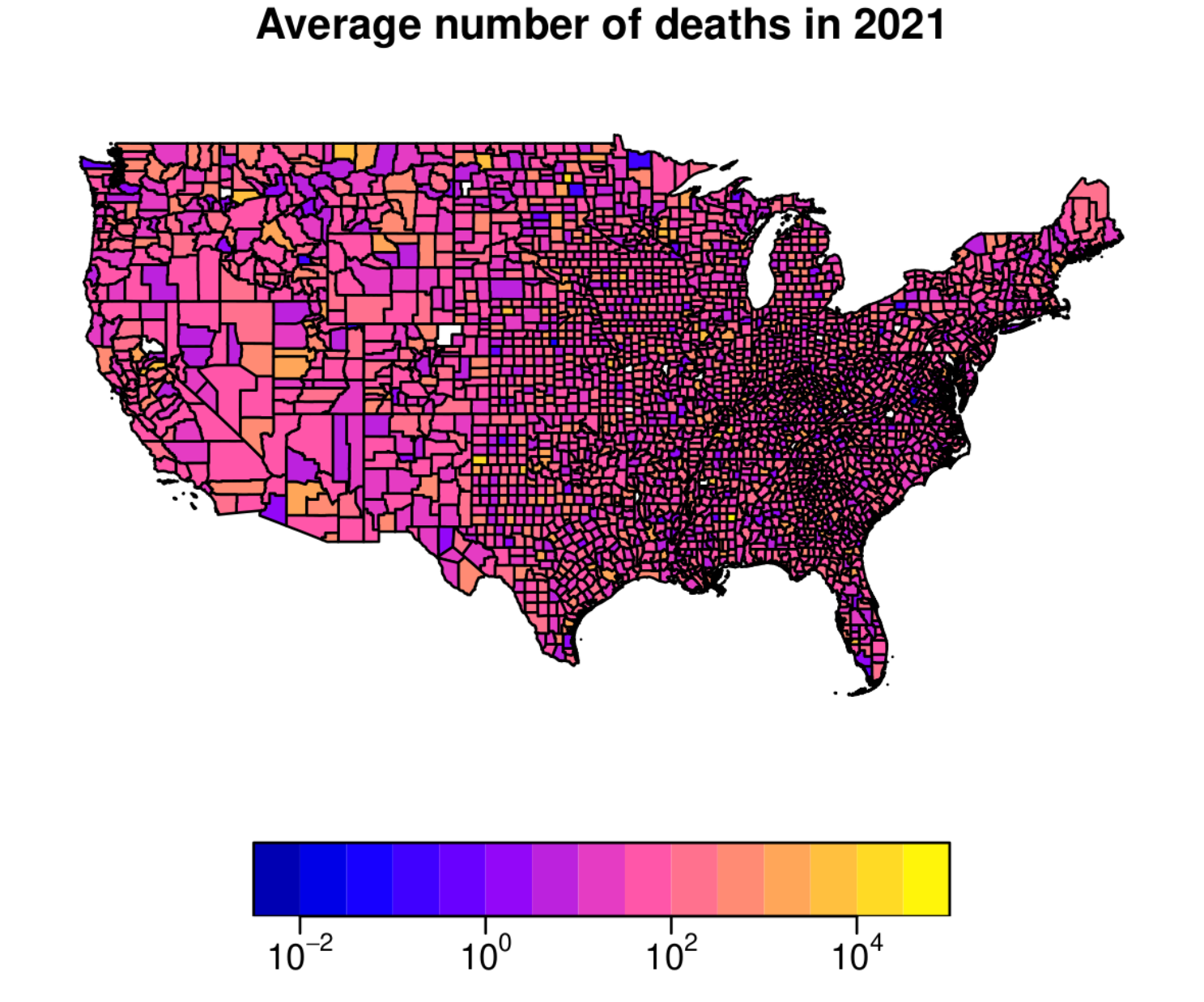}
\includegraphics[width=8.9cm]{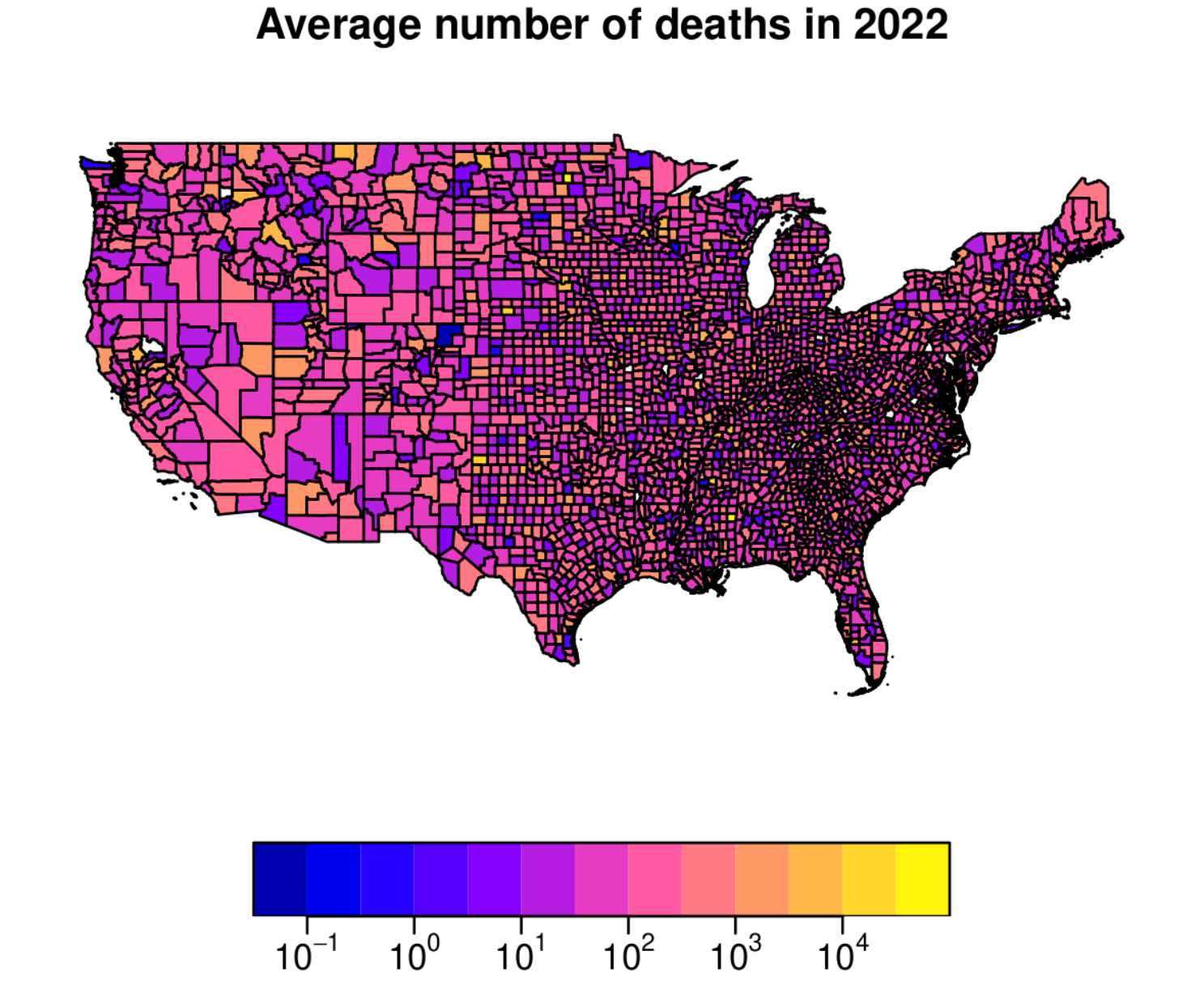}
\caption{\small{Choropleth maps depicting the average confirmed cases (top panels) and the average number of deaths (bottom panels) for U.S. cities in the years 2021 (left panels) and 2022 (right panels).}}\label{fig:Fig_2}
\end{figure}

In Figure~\ref{fig:Fig_2}, we present choropleth maps illustrating the average confirmed COVID-19 cases and the average number of deaths for U.S. cities. The clusters of similar colors observed in these maps indicate potential spatial dependence in both confirmed COVID-19 cases and the average number of deaths. This suggests that neighboring cities tend to share similar characteristics regarding the average confirmed cases and the average number of deaths. As shown in Figures~\ref{fig:Fig_1} and~\ref{fig:Fig_2}, some cities demonstrate exceptionally high average numbers of deaths and confirmed cases compared to others, suggesting the presence of outliers in both the response variable (average number of deaths) and the predictor variable (confirmed cases).

We delve into the spatial correlation and outlier analysis of the U.S. COVID-19 data set through the computation of the local Moran's I statistic. We begin by constructing the elements of the weight matrix \(\bm{W} = (w_{ii^{\prime}})_{1 \leq i,i^{\prime} \leq n}\) based on the inverse distance between two cities, i.e., \(w_{ii^{\prime}} = \frac{1/d_{ii^{\prime}}}{\sum_{l=1}^n d_{il}}\), where \(d_{ii^{\prime}}\) is determined using the great circle distance between cities \(i\) and \(i^{\prime}\). This distance is calculated using the Haversine formula on the internal points of the counties: \(d_{ii^{\prime}} = Rc\), where \(c = 2 \times \atantwo (\sqrt{a}, \sqrt{1-a})\) with \(a = \sin^2(\Delta u / 2) + \cos(u_1) \cos(u_2) \sin^2(\Delta v / 2)\). Here, \(u\) represents the latitude, \(v\) represents the longitude, and \(R\) represents the Earth’s radius (mean radius = 6,371 km).

Local indicators of spatial association, such as the local Moran's I statistic, aim to provide insight into the extent of significant spatial clustering of similar values around each observation \citep{Ansellin1995}. For the \(i^\text{th}\) location, the local Moran's I statistic is defined as follows:
\begin{equation*}
I_i = \frac{n (Y_i - \overline{Y})}{\sum_{i^{\prime} = 1}^n (Y_{i^{\prime}} - \overline{Y})^2} \sum_{i^{\prime} = 1}^n w_{i i^{\prime}} (Y_{i^{\prime}} - \overline{Y}).
\end{equation*}
In practice, local Moran's I values are often used for spatial mapping, offering insights into areas with notably strong or weak associations with neighboring regions. A high \(I_i\) value indicates that the area is part of a cluster of observations, which can be high, low, or moderate. Conversely, a low \(I_i\) suggests that the area is surrounded by regions with disparate values. This information collectively enables the classification of significant locations into distinct spatial clusters, such as High-High, Low-Low, High-Low, and Low-High.

\begin{figure}[!htb]
\centering
\includegraphics[width=9.2cm]{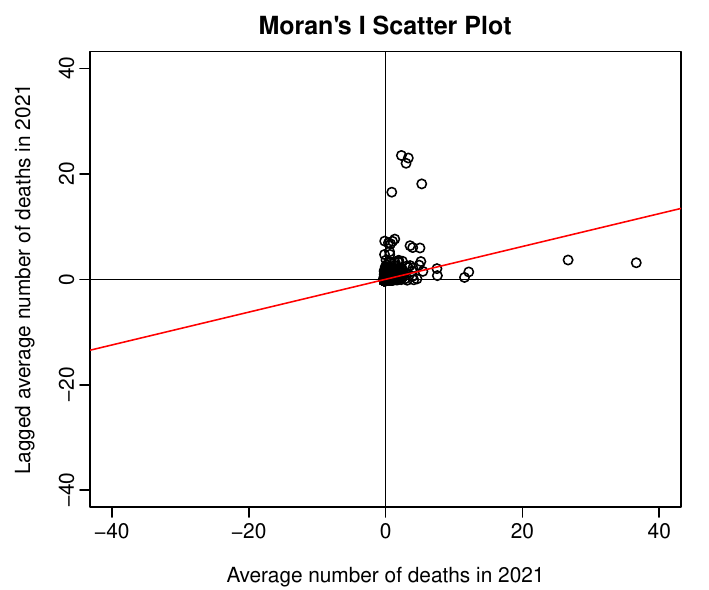}
\caption{\small{Moran's I scatter plot of local spatial autocorrelation of the average number of deaths in the U.S. in 2021}.}\label{fig:Fig_3}
\end{figure}

The scatter plot displaying the computed \(I_i\) values is presented in Figure~\ref{fig:Fig_3}. Examining this plot reveals that \(I_i\) values predominantly fall within the first quadrant of the \(X-Y\) plot, indicating areas with high values surrounded by similarly high values. In essence, Figure~\ref{fig:Fig_3} illustrates a concentration of values in regions exhibiting positive autocorrelation, thereby demonstrating a positive spatial autocorrelation in the average number of COVID-19-related deaths across the U.S.

Figure~\ref{fig:Fig_3} highlights certain observations as outliers. These initial analyses provide a compelling rationale for applying our proposed RFPC and RFPLS-based SSoFRMs, which offer a robust and effective approach to exploring the dynamics of COVID-19 prevalence in the U.S., while accounting for spatial dependencies.

With the COVID-19 data set, we consider the following SSoFRM:
\begin{equation*}
Y = \beta_0 \bm{1}_n + \rho \bm{W} Y + \int_{t=1}^{365} \X(t) \beta(t) dt + \epsilon.
\end{equation*}
We apply FPC and FPLS methods, along with their robust counterparts, to estimate the SSoFRM using the 2021 COVID-19 data set. To evaluate the in-sample performance, we compute metrics such as the mean squared error (MSE) and the coefficient of determination (\(R^2\)), as well as their trimmed versions:
\begin{equation*}
\text{MSE} = \sum_{i=1}^{n^*} (Y_i - \widehat{Y}_i)^2, \quad R^2 = 1 - \frac{\sum_{i=1}^{n^*}(Y_i - \widehat{Y}_i)^2}{\sum_{i=1}^{n^*}(Y_i - \overline{Y})^2},
\end{equation*}
where \(n^*\) denotes the number of samples with no outliers, so \(n^* = n\) when the trimming percentage is \(\iota = 0\). The trimmed observations are identified based on the squared residuals, i.e., \(\widehat{\epsilon}_i = Y_i - \widehat{Y}_i\). Furthermore, we utilize the fitted models to predict the average number of deaths for the year 2022 based on the number of confirmed cases. This allows us to evaluate out-of-sample performance. Metrics such as the mean squared prediction error (MSPE) and coefficient of determination for the test sample (\(R_p^2\)), similar to MSE and \(R^2\) defined earlier, are computed to assess and compare the predictive accuracy of the methods.

The estimated values of \(\rho\) and \(\sigma\) from the 2021 data are presented in Table~\ref{tab:tab_3}. All methods estimate spatial autocorrelation at approximately 0.3, consistent with the observations in Figure~\ref{fig:Fig_3}. Table~\ref{tab:tab_3} demonstrates that the proposed robust methods yield significantly smaller \(\widehat{\sigma}\) compared to their non-robust counterparts. This difference is attributed to outlier observations in both the response and predictor variables across neighboring counties. 
\begin{table}[!htb]
\centering
\tabcolsep 0.5in
\caption{Estimated $\rho$ and $\sigma$ values from the COVID-19 data for the year 2021.}\label{tab:tab_3}
\begin{tabular}{@{}lcccc@{}} 
\toprule
{Statistic} & FPC & FPLS & RFPC & RFPLS \\
\cmidrule(l){2-5}
$\widehat{\rho}$ & 0.301 & 0.301 & 0.294 & 0.294 \\
$\widehat{\sigma}$ & 312.536 & 312.793 & 49.981 & 50.031 \\
\bottomrule
\end{tabular}
\end{table}

\begin{figure}[!htb]
\centering
\includegraphics[width=9.5cm]{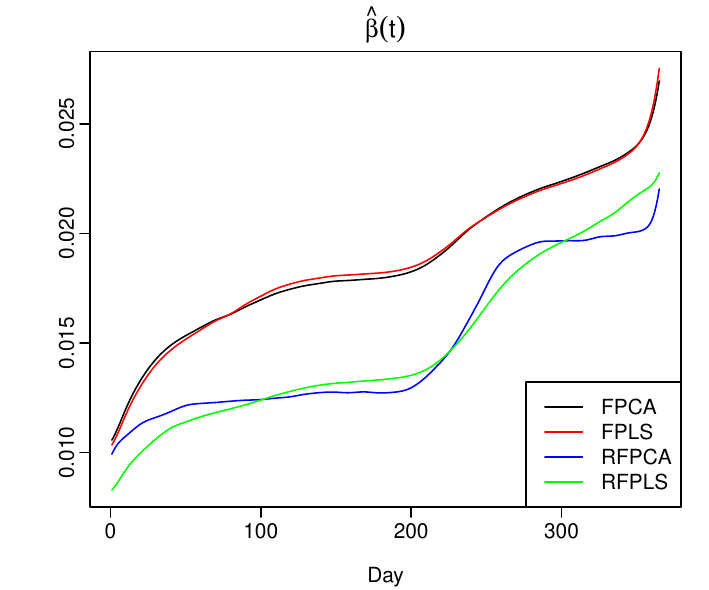}
\caption{\small{The estimated regression coefficient functions for the U.S. COVID-19 data set in the year 2021}.}\label{fig:Fig_4}
\end{figure}

The estimated regression coefficient functions are illustrated in Figure~\ref{fig:Fig_4}. These functions consistently depict an increasing trend over time, indicating that the number of confirmed cases positively influences the average number of deaths.

This relationship is well supported by the data presented in Figure~\ref{fig:Fig_1}, which illustrates a general rise in the number of confirmed cases across all U.S. cities, correlating with an increase in deaths. Moreover, Figure~\ref{fig:Fig_1} highlights that the regression coefficient functions derived from non-robust methods exert a stronger influence on the average number of deaths compared to the proposed robust methods, particularly noticeable in the last 65-70 days of 2021. This discrepancy is attributed to a significant surge in confirmed cases during this period and the presence of outlier curves in the predictor variable. Non-robust methods are more sensitive to these outliers, resulting in larger values for the final period of 2021. In contrast, the proposed robust methods mitigate the impact of outlier observations, leading to comparatively smaller values for these days.

The computed MSE, \(R^2\), MSPE, and \(R_p^2\) values are displayed in Table~\ref{tab:tab_4}. For the entire data set (\(\iota = 0\)), non-robust methods generally provide better in-sample predictions (i.e., smaller MSE and higher \(R^2\) values) compared to the proposed robust methods. This aligns with expectations, as maximum-likelihood-type estimates typically excel in providing accurate in-sample inference for the entire data set, outperforming robust methods.
\begin{table}[!htb]
\centering
\tabcolsep 0.107in
\caption{Computed MSE ($10^4$), $R^2$, MSPE ($10^4$), and $R_p^2$ values for the U.S. COVID-19 data set. The results are obtained when the trimming percentage (TP) is 0\% (i.e., using full data set), 5\% and 10\%.}\label{tab:tab_4} 
\begin{tabular}{@{}lcccccccccccc@{}} 
\toprule
{Method} & \multicolumn{4}{c}{TP = 0\%} & \multicolumn{4}{c}{TP = 5\%} & \multicolumn{4}{c}{TP = 10\%} \\
\cmidrule(l){2-13}
& MSE & $R^2$ & MSPE & $R_p^2$ & MSE & $R^2$ & MSPE & $R_p^2$ & MSE & $R^2$ & MSPE & $R_p^2$ \\
\cmidrule(l){2-13}
FPC & \textbf{0.847} & \textbf{0.872} & 81.117 & \textbf{0.932} & 0.370 & \textbf{0.963} & 5.957 & 0.891 & 0.211 & \textbf{0.979} & 2.853 & 0.901 \\
FPLS & 9.863 & \textbf{0.872} & 81.359 & \textbf{0.932} & 0.370 & \textbf{0.963} & 6.007 & 0.891 & 0.211 & \textbf{0.979} & 2.871 & 0.901 \\
RFPC & 12.540 & 0.871 & 29.382 & \textbf{0.932} & \textbf{0.212} & \textbf{0.963} & 2.626 & \textbf{0.932} & \textbf{0.109} & 0.973 & 1.412 & \textbf{0.936} \\
RFPLS & 16.843 & 0.851 & \textbf{29.067} & \textbf{0.932} & 4.371 & 0.925 & \textbf{2.595} & \textbf{0.932} & 4.244 & 0.928 & \textbf{1.394} & \textbf{0.936} \\
\bottomrule
\end{tabular}
\end{table}

Conversely, the proposed robust methods demonstrate improved out-of-sample predictions relative to non-robust methods in this scenario. We attribute this to the presence of larger outliers in the 2021 data set compared to 2022, which biases the estimates obtained by non-robust methods and compromises their predictive reliability for the entire data set, that is 2022. In contrast, the robust estimates provided by our methods lead to more dependable predictions across both in-sample and out-of-sample scenarios.

For \(\iota \neq 0\), where data is trimmed, the proposed robust methods consistently show better in-sample and out-of-sample predictions compared to non-robust methods. Specifically, the RFPC-based SSoFRM yields significantly smaller in-sample prediction errors, with lower MSE and higher \(R^2\) values. In contrast, the RFPLS-based SSoFRM produces the most accurate out-of-sample predictions among its counterparts. While RFPLS results in larger in-sample prediction errors compared to RFPC, it achieves better out-of-sample performance, with smaller MSPE and higher \(R^2\)) values than RFPC.

\section{Conclusion}\label{sec:5}

In this study, we have introduced robust estimation methods for the SSoFRM using FPC and FPLS techniques. By mitigating the vulnerability of traditional maximum likelihood estimators to outliers, our RFPC and RFPLS approaches significantly improve parameter estimation accuracy. These methods achieve this by projecting the infinite-dimensional functional predictor onto a finite-dimensional space and employing M-estimation techniques, effectively reducing the influence of leverage points and vertical outliers. 

Simulation studies and empirical data analysis validate that the proposed RFPC and RFPLS methods offer enhanced robustness compared to conventional approaches. This robustness makes them well-suited for modeling complex phenomena such as COVID-19 dynamics, where outlier resilience is crucial for accurate inference and prediction.

Our research contributes to the broader understanding and application of spatial functional linear regression models in fields characterized by spatially correlated data. The robust methods developed in this paper pave the way for more reliable and accurate modeling in various domains, including demography, economics, environmental sciences, agronomy, and mining. Our proposed methodology opens several possibilities for further research and enhancement, as detailed below:
\begin{asparaenum}
\item[1)] The current study utilizes a single functional predictor variable in the SSoFRM. Future work can expand our method to incorporate multiple functional predictors, potentially enhancing estimation and predictive outcomes.
\item[2)] While our study focuses solely on functional predictors, it is feasible to include a combination of functional and scalar predictors, akin to a partial functional linear model, to account for variation in scalar responses. Our method can be adapted to integrate scalar predictors seamlessly.
\item[3)] The performance of our estimation method can be enhanced by incorporating a penalization term to regulate the smoothness of the functional variables. Developing a penalized version of our approach could lead to superior estimation results in scalar-on-function regression.
\item[4)] Incorporating modern dimension reduction techniques, such as projection pursuit \citep{Friedman1974} and invariant coordinate selection \citep{Tyler2009}, into our proposed methods could significantly enhance their estimation and predictive performance.
\end{asparaenum}

\section*{Acknowledgment}

We thank two reviewers for their thorough reading of our manuscript and their insightful suggestions and comments, which have significantly contributed to the improvement of our article. This research was supported by the Scientific and Technological Research Council of Turkey (TUBITAK) (under grant no.~124F096), the Australian Research Council Discovery Project (grant no.~DP230102250), and the Australian Research Council Future Fellowship (grant no.~FT240100338).

\appendix

\section{Conditions}

We present an analysis of the stability and impact characteristics of the RFPC-based estimators outlined in Sections~\ref{sec:3.1} and~\ref{sec:3.3}. Additionally, we delve into elucidating the asymptotic distribution of these estimators. For the SSoFRM, the RFPC-based estimate of the regression coefficient function $\widehat{\beta}(t)$ is derived using the M-estimator within the RFPC eigenspace spanned by the robust eigenfunctions of the covariance operator. Additionally, the estimates $[\widehat{\beta}_0, \widehat{\sigma}, \widehat{\rho}]$ are obtained using the M-estimator introduced by \cite{Tho2023}. Despite potential dissimilarities with estimates obtained via the maximum likelihood estimator in the classical FPC eigenspace spanned by the eigenfunctions, these estimates exhibit asymptotic consistency under certain regularity conditions on the processes $Y$, $\X(t)$, the matrix of basis expansion coefficients $\bm{A}$, and the spatial weight matrix $\bm{W}$, as outlined in the following theorem and corollary.

Commencing with a $p \times q$-dimensional matrix $\bm{X}$, let $\vert \bm{X} \vert_{\infty}$ = $\max_{1 \leq i \leq p, 1 \leq j \leq q} \vert x_{i,j} \vert$, $\Vert \bm{X} \Vert_{\infty} = \max_{1 \leq i \leq p} \sum_{j=1}^q \vert x_{ij} \vert$, and $\Vert \bm{X} \Vert_1 = \max_{1 \leq j \leq q} \sum_{i=1}^p \vert x_{ij} \vert$. Subsequently, we consider the matrix formulation of the approximate SSoFRM:
\begin{equation}\label{eq:fm}
Y = \rho \bm{W} Y + \bm{Z} \bm{\theta} + \epsilon,
\end{equation}
where $\bm{Z} = [\bm{1}_n, \bm{A}]^\top$ and $\bm{\theta} = [\beta_0, \bm{\beta}^\top]^\top$ with $\bm{\beta} = [\widehat{\beta}_1, \ldots, \widehat{\beta}_K]$. Let $\bm{\Theta} = [\bm{\theta}, \sigma, \rho]^\top$ denote the parameter vector for Model~\eqref{eq:fm} and let $\bm{\Theta}_0 = [\bm{\theta}_0, \sigma_0, \rho_0]^\top$ denotes the true value of the model parameters. In addition, let $\mathcal{\bm{V}}(\bm{\Theta}_0) = n^{-1} \text{E}[\eta_R(\bm{\Theta}_0; Y) \eta_R(\bm{\Theta}_0; Y)^\top]$ and $\mathcal{\bm{V}}^*(\bm{\Theta}_0) = -n^{-1} \text{E}[\partial \eta_R(\bm{\Theta}_0; Y) / \partial \bm{\Theta}^\top]$. To establish the reliability of the estimators, we outline the requisite conditions.

\begin{itemize}
\item[$C_1$.] The functional process $\X(t)$ possesses a finite-dimensional Karhunen-Lo\`{e}ve decomposition, represented as $\X(t) = \sum_{k=1}^K a_k \phi_k(t)$, with eigenvalues $\lambda_1 > \ldots > \lambda_K > 0$.
\item[$C_2$.] The regression coefficient function $\beta(t)$ resides in a linear subspace spanned by $\{\phi_1(t), \ldots, \phi_K(t) \} \in \mathcal{L}^2(\mathcal{I})$.
\item[$C_3$.] The random variables $\{a_1, \ldots, a_K \}$ linked with $\{\phi_1(t), \ldots, \phi_K(t) \}$ are absolutely continuous, exhibiting a joint density $g(x)$ satisfying $g(x) = \iota(\Vert x \Vert_E)$ for $x \in \mathbb{R}^K$ for some measurable function $\iota: \mathbb{R} \rightarrow \mathbb{R}_{+}$, where $\Vert \cdot \Vert_E$ denotes the Euclidean norm. 
\item[$C_4$.] The functional process $\X$ has finite fourth moments, i.e., $\text{E}[\Vert \X \Vert^4] < \infty$, where $\Vert \X \Vert^4$ refers to the fourth power of the norm of the functional process $\X$ in the space $\mathcal{L}^2(\mathcal{I})$.
\item[$C_5$.] The elements of the model matrix $\bm{Z}$ are uniformly bounded constants for all $n \geq 1$, specifically, $\sup_{n \geq 1} \Vert \bm{Z} \Vert_{\infty} < \infty$.
\item[$C_6$.] The spatial weight matrix $\bm{W}$ takes the form $w_{i i^{\prime}} = \frac{ m(d_{i i^{\prime}})}{\sum_{j^{\prime}=1}^n m(d_{i j^{\prime}})}$, and the true correlation parameter $\rho_0$ satisfies $\rho_0 \in (- \vert \lambda_{\min}(\bm{W}) \vert^{-1}, 1)$ for such row-normalized $\bm{W}$.
\item[$C_7$.] The row sums and column sums of matrices $\bm{W}$ and $(\mathbb{I}_n - \rho \bm{W})^{-1}$ are uniformly bounded for all $n \geq 1$ and $\rho \in (- \vert \lambda_{\min}(\bm{W}) \vert^{-1}, 1)$, that is, $\sup_{n \geq 1} \Vert \bm{W} \Vert_1 < \infty$, $\sup_{n \geq 1} \Vert \bm{W} \Vert_{\infty} < \infty$, $\sup_{n \geq 1, \rho \in (- \vert \lambda_{\min}(\bm{W}) \vert^{-1}, 1)} \Vert (\mathbb{I}_n - \rho \bm{W})^{-1} \Vert_1 < \infty$, and $\sup_{n \geq 1, \rho \in (- \vert \lambda_{\min}(\bm{W}) \vert^{-1}, 1)} \Vert (\mathbb{I}_n - \rho \bm{W})^{-1} \Vert_{\infty} < \infty$.
\item[$C_8$.] The matrices $\mathcal{\bm{V}}(\bm{\Theta}_0) \rightarrow \widetilde{\mathcal{\bm{V}}}(\bm{\Theta}_0)$ and $\mathcal{\bm{V}}^*(\bm{\Theta}_0) \rightarrow \widetilde{\mathcal{\bm{V}}}^*(\bm{\Theta}_0)$ as $n \rightarrow \infty$, where $\widetilde{\mathcal{\bm{V}}}(\bm{\Theta}_0)$ and $\widetilde{\mathcal{\bm{V}}}^*(\bm{\Theta}_0)$ are positive definite matrices.
\end{itemize}

Conditions $C_1$ and $C_2$ are fundamental to ensure that the infinite-dimensional SSoFRM can be effectively represented with a finite number of truncation constants denoted as $K$. The significance of Condition $C_3$ lies in establishing the Fisher-consistency of the RFPC estimator. The satisfaction of Conditions $C_1$ and $C_3$ is contingent upon the validity of $C_4$. In order to establish the Fisher-consistency of the M-estimators $[\widehat{\beta}_0, \widehat{\sigma}, \widehat{\rho}]$, adherence to Conditions $C_5-C_8$ is imperative. Condition $C_5$ assumes the boundedness of random variables associated with the functional predictor, simplifying the consideration of unbounded error terms on the SSoFRM within the finite-dimensional space of the RFPC basis expansion coefficients. Condition $C_6$ ensures the non-singularity of the matrix $(\mathbb{I}_n - \rho_0 \bm{W})$ and the existence of equilibrium. Condition $C_7$ holds when $\rho_0 = \rho$, limiting spatial dependence. Together with Condition $C_8$, these conditions are necessary to establish a central limit theorem for the estimators.
\section{Consistency}

\begin{theorem}\label{teo:1}
Under conditions $C_1-C_8$, the RFPC-based estimators $[\widehat{\beta}_0, \widehat{\beta}(t), \widehat{\sigma}, \widehat{\rho}]$ are Fisher-consistent.
\end{theorem}

\begin{proof}[Proof of Theorem~\ref{teo:1}]
In the proof, we utilize techniques similar to those explained by \cite{kalogridis2019} and \cite{Tho2023}. Let $\mathbb{P}$ denote the image measure of $\X$, represented by $\mathbb{P}(U) = P(\X \in U)$ for a Borel set $U$. Consequently, the distribution function of $\X$ is defined as follows:
\begin{equation*}
\mathcal{F}(\tau_1, \ldots, \tau_K) := \mathcal{P}(a_1 \leq \tau_1, \ldots, a_K \leq \tau_K).
\end{equation*}
Next, the functional of the proposed robust estimator for the regression coefficient function $\widehat{\beta}(t)$ is defined by:
\begin{equation*}
\widehat{\beta}(\mathcal{F}) = \sum_{k=1}^K \widehat{\beta}_k(\mathcal{F}) \widehat{\phi}_k(\mathcal{F})(t).
\end{equation*}
The term ``Fisher-consistent'' is attributed to the functional $\widehat{\beta}(t)$ under the prerequisite that $\widehat{\beta}(\mathcal{F}) = \beta(t)$ holds for all $t \in \mathcal{I}$. This implies that $\widehat{\beta}_k(\mathcal{F}) = \beta_k$ and $\widehat{\phi}_k(\mathcal{F})(t) = \phi_k(t)$. Put simply, the functional $\widehat{\beta}(t)$ is deemed Fisher-consistent if the M-estimator of location for functional data, the RFPC estimator, and the M-estimator introduced by \cite{Tho2023} all demonstrate Fisher-consistency.

In line with the findings of \cite{kalogridis2019}, satisfaction of condition $C_3$ proves to be adequate for ensuring the Fisher-consistencies of both the M-estimator of location for functional data and the RFPC estimator. Subsequently, under condition $C_1$, the representation $\X(t) = \sum_{k=1}^K a_k \widehat{\phi}_k(\mathcal{F})(t)$ is established. For the SSoFRM, by leveraging conditions $C_1$ and $C_2$ in conjunction with the orthonormal properties inherent in $\widehat{\phi}_k(\mathcal{F})(t)$, we derive:
\begin{align}
Y &= \beta_0 \bm{1}_n + \rho \bm{W} Y + \sum_{k = 1}^K a_k \widehat{\phi}_k(\mathcal{F})(t) \widehat{\phi}^\top_k(\mathcal{F})(t) \widehat{\beta}_k(\mathcal{F}) + \epsilon \nonumber \\
& = \beta_0 \bm{1}_n + \rho \bm{W} Y + \bm{A}^\top \widehat{\bm{\beta}}(\mathcal{F}) + \epsilon. \label{eq:fm1}
\end{align}

Consider the vectors $\bm{Z} = [\bm{1}_n, \bm{A}]^\top$ and $\bm{\theta} = [\beta_0, \bm{\beta}^\top(\mathcal{F})]^\top$, where $\bm{\beta}^\top(\mathcal{F}) = [\widehat{\beta}_1(\mathcal{F}), \ldots, \widehat{\beta}_K(\mathcal{F})]^\top$ represents the model parameters. The matrix form for Model~\eqref{eq:fm1}, or equivalently Model~\eqref{eq:fm}, is established. Leveraging Lemmas 1-8 in the supplementary file and Theorem 1 from \cite{Tho2023}, subject to the conditions stipulated in $C_5-C_8$, we observe that $\frac{1}{\sqrt{n}} \eta_R(\bm{\Theta}_0; Y) \xrightarrow[]{d} \text{N}(\bm{0}, \widetilde{\mathcal{\bm{V}}}(\bm{\Theta}_0))$. This result implies that $\text{E}[\eta_R(\bm{\Theta}_0, Y)] = \bm{0}$, establishing the Fisher-consistency of the parameter matrix $\bm{\Theta} = [\bm{\theta}, \sigma, \rho]^\top$, where $\bm{\theta} = [\beta_0, \bm{\beta}^\top]^\top$. Considering the collective Fisher-consistencies of the M-estimator of location for functional data, RFPCA estimator, and $\bm{\Theta}$, the RFPC-based M-estimators obtained through Algorithm~\ref{alg:estalg} exhibit Fisher-consistency, and we affirm that $\widehat{\beta}(\mathcal{F}) = \beta(t)~ \forall t \in \mathcal{I}$.
\end{proof}

\begin{corollary}\label{cor:1}
If $\lbrace (Y_1, \X_1(t)), \ldots, (Y_n, \X_n(t)) \rbrace$ are i.i.d. samples with cumulative distributions $(\mathcal{F}_Y, \mathcal{F}_{\X})$. Then, under the conditions $C_1-C_8$, the estimators $[\widehat{\beta}_0, \widehat{\beta}(t), \widehat{\sigma}, \widehat{\rho}]$ are consistent.
\end{corollary}

\begin{proof}[Proof of Corollary~\ref{cor:1}]
For an i.i.d. random sample denoted as $\lbrace (Y_1, \X_1(t)), \ldots, (Y_n, \X_n(t)) \rbrace$ characterized by distribution functions $(\mathcal{F}_Y, \mathcal{F}_{\X})$, the assumption in condition $C_1$ implies that the definitions of $\X_i(t)$ involve a finite number of eigenfunctions and their corresponding random variables. Utilizing the Glivenko-Cantelli theorem \citep[see, for instance,][]{polard1984}, which establishes the uniform convergence of empirical distribution functions to their population counterparts, it can be demonstrated in the context of Theorem~\ref{teo:1} and adherence to conditions $C_1-C_8$ that $\widehat{\beta}(t)$ achieves asymptotic consistency. This assertion is grounded in the understanding that Fisher-consistency is equivalent to asymptotic consistency, given the point-wise convergence of empirical distribution functions of random variables to their population distribution functions.
\end{proof}

\section{Influence function}

We will evaluate the robustness of our proposed estimator to outliers by employing the influence function method introduced by \cite{Hampel}. This method quantifies the asymptotic bias rate of an estimator when subjected to minimal contamination within the distribution. An estimator with a bounded influence function is considered robust against extreme outliers. Let $(\Y, \X^{\top})^{\top} \in \mathbb{R}^p$ be a random variable with distribution $F_{\bm{\Theta}}=(G_0 \times \mathbb{P}_x)$, where $\bm{\Theta} \in \bm{\Theta}_0 \subset \mathbb{R}^m$. Assume $T$ to be an estimating functional of $\bm{\Theta}$ that is Fisher consistent, meaning $T(F_{\bm{\Theta}}) = \bm{\Theta}$. The influence function for $T$ is given by
\begin{equation*}
	\text{IF}(\bm{s}, T, F_{\bm{\Theta}}) = \lim_{\epsilon \downarrow 0}\frac{T(F_{{\bm{\Theta}}, \epsilon}) -T(F_{\bm{\Theta}}) }{\epsilon} = \frac{\partial}{\partial \epsilon} T(F_{\bm{\Theta}})\Big|_{\epsilon=0} ,
\end{equation*}
where $F_{{\bm{\Theta}}, \epsilon} = (1-\epsilon) F_{\bm{\Theta}} + \epsilon \delta_{\bm{s}}$ represents the contaminated distribution and $\delta_{\bm{s}}$ is the point-mass distribution at $\bm{s}=(\Y_0, \X_0) \in \mathbb{R}^K$. Hence, the influence function is the Gateaux derivative of the functional $T$, defined on the space of finite signed measures, in the direction $\delta_{\bm{s}} - F_{\bm{\Theta}}$. The following theorem discusses the influence function of our estimator.

\begin{theorem}\label{the:if}
Let $\widehat{\bm{\Theta}} = (\widehat{\beta}_0, \widehat{\beta}^{\top}, \widehat{\sigma}, \widehat{\rho})^{\top}$ denote the SSoFRM estimator, which is Fisher consistent, and consider the contaminated distribution $F_{{\bm{\Theta}}, \epsilon} = (1-\epsilon) F_{\bm{\Theta}} + \epsilon \delta_{\bm{s}}$ where $\bm{s}=(\Y_0, \X_0) \in \mathbb{R}^K$. The influence function of the regression coefficients $\beta(t)$ is bounded and continuous in $Y$. However, the influence function may be unbounded in $\X$ particularly when there is a large discrepancy between the eigenfunction scores corresponding to different eigenvalues of the covariance operator of  $\X(t)$. Despite this, the influence function of $\beta(t)$ is impacted primarily by good leverage points, and only such points can significantly affect the estimators.
\end{theorem}

\begin{proof}[Proof of Theorem~\ref{the:if}]
The  SSoFRM estimator $\widehat{\bm{\Theta}} = (\widehat{\beta}_0, \widehat{\beta}^{\top}, \widehat{\sigma}, \widehat{\rho})^{\top}$ is the solution of the following estimating equation: $\eta_R(\bm{\Theta}; Y) = 0$. As the SSoFRM estimator is Fisher consistent, we have:
\begin{equation}
 \int_\X \int_\Y \eta_R(\bm{\Theta}; Y) d F_{\bm{\Theta}} = 0.
 \label{est_eqn}
\end{equation}
Let $\bm{\Theta}_\epsilon$ be the value of the parameter under the contaminated distribution $F_{{\bm{\Theta}}, \epsilon} = (1-\epsilon) F_{\bm{\Theta}} + \epsilon \delta_{\bm{s}}$, where $\bm{s}=(\Y_0, \X_0)$. Then,~\eqref{est_eqn} can be expressed as
\begin{equation}\label{est_eqn2}
 \int_\X \int_\Y \eta_R(\bm{\Theta}_\epsilon; Y) d F_{{\bm{\Theta}}, \epsilon} = 0.
\end{equation}
Differentiating~\eqref{est_eqn2} with respect to $\epsilon$, we get
\begin{small}
\begin{equation*}
   \int_\X \int_\Y \left\{\frac{\partial}{\partial {\bm{\Theta}}} \eta_R(\bm{\Theta}; Y) \right\}\Bigg|_{{\bm{\Theta}} = {\bm{\Theta}}_\epsilon} \frac{\partial {\bm{\Theta}}_\epsilon}{\partial \epsilon}  d F_{{\bm{\Theta}}, \epsilon}  
  + \int_\X \int_\Y \left\{\frac{\partial}{\partial {\rm vec}(\bm{Z})} \eta_R(\bm{\Theta}; Y)  \frac{\partial {\rm vec}(\bm{Z})}{\partial \epsilon} \right\}\Bigg|_{{\bm{\Theta}} = {\bm{\Theta}}_\epsilon} d F_{{\bm{\Theta}}, \epsilon} 
  + \int_\X \int_\Y \eta_R(\bm{\Theta}_\epsilon; Y) (d F_{\bm{\Theta}} + d \delta_{\bm{s}})  = 0.
\end{equation*}
\end{small}
Therefore, we have
\begin{small}
\begin{align*}
\frac{\partial {\bm{\Theta}}_\epsilon}{\partial \epsilon} = - \Big[ 
 \int_\X \int_\Y \left\{\frac{\partial}{\partial {\bm{\Theta}}} \eta_R(\bm{\Theta}; Y) \right\}\Bigg|_{{\bm{\Theta}} = {\bm{\Theta}}_\epsilon}  d F_{{\bm{\Theta}}, \epsilon} \Big]^{-1}\times \Bigg[
\int_\X  \int_\Y \left\{\frac{\partial}{\partial {\rm vec}(\bm{Z})} \eta_R(\bm{\Theta}; Y)   \right\}\Bigg|_{{\bm{\Theta}} = {\bm{\Theta}}_\epsilon} \frac{\partial {\rm vec}(\bm{Z})}{\partial \epsilon} d F_{{\bm{\Theta}}, \epsilon}  \\
+ \int_\X \int_\Y \eta_R(\bm{\Theta}_\epsilon; Y) (d F_{\bm{\Theta}} + d \delta_{\bm{s}} )
\Bigg].
\end{align*}
\end{small}
Now, evaluating both side at $\epsilon=0$, we get the influence function:
\begin{equation}
\text{IF}(\bm{s}, \bm{\Theta}, F_{\bm{\Theta}}) = - \Big[ 
\int_\X \int_\Y \frac{\partial}{\partial {\bm{\Theta}}} \eta_R(\bm{\Theta}; Y)   d F_{\bm{\Theta}}  
\Big]^{-1} \times \Bigg[
\int_\X  \int_\Y \frac{\partial}{\partial {\rm vec}(\bm{Z})} \eta_R(\bm{\Theta}; Y)   \frac{\partial {\rm vec}(\bm{Z})}{\partial \epsilon} \Bigg|_{\epsilon = 0} d F_{\bm{\Theta}}  +  \eta_R(\bm{\Theta}; \Y_0)  \Big|_{\X = \X_0}
\Bigg]. \label{if}
\end{equation}
Define $F_{\bm{\Theta}}(\X)$ and $F_{\bm{\Theta}}(\Y)$ as the marginal distribution functions of $\X$ and $\Y$, respectively. Following \cite{Tho2023}, we find that 
\begin{align}
	-\int_\Y \frac{\partial}{\partial {\bm{\Theta}}} \eta_R(\bm{\Theta}; Y)   d F_{\bm{\Theta}}(\Y) = \bm{B} (\bm{\Theta})= \begin{pmatrix}
		b_{\theta \theta}(\bm{\Theta}) & b_{\theta \sigma}(\bm{\Theta}) & b_{\theta \rho}(\bm{\Theta})\\
		b_{\sigma \theta}(\bm{\Theta}) & b_{\sigma \sigma}(\bm{\Theta}) & b_{\sigma \rho}(\bm{\Theta})\\
		b_{\rho \theta}(\bm{\Theta}) & b_{\rho \sigma}(\bm{\Theta}) &b_{\rho \rho}(\bm{\Theta})
	\end{pmatrix},
	\label{b_mat}
\end{align}
where
\begin{align*}
b_{\theta \theta}(\bm{\Theta}) & = \frac{1}{\sigma}\{2\Phi(c_1) - 1\} \bm{Z}^{\top} \bm{Z},\\
b_{\theta \sigma}(\bm{\Theta}) & = \bm{0}_{K+1},\\
b_{\theta \rho}(\bm{\Theta}) & =  \frac{1}{\sigma}\{2\Phi(c_1) - 1\} \bm{Z}^{\top} \bm{G}(\rho) \bm{Z} \bm{\theta},\\
b_{\sigma \theta}(\bm{\Theta}) & = \bm{0}_{K+1}^{\top}, \\
b_{\sigma \sigma}(\bm{\Theta}) & =   \frac{2}{\sigma}\{2\Phi(c_2) - 1 -2c_2 \Phi'(c_2)\}, \\
b_{\sigma \rho}(\bm{\Theta}) & = 2 tr\{\bm{G}(\rho)\} \{2\Phi(c_2) - 1 -2c_2 \Phi'(c_2) \}, \\
b_{\rho \theta}(\bm{\Theta}) & = \frac{1}{\sigma^2}\{2\Phi(c_3) - 1\} \{ \bm{G}(\rho) \bm{Z} \bm{\theta}\}^{\top} \bm{Z},\\
b_{\rho \sigma}(\bm{\Theta}) & =  \frac{2}{\sigma} tr\{\bm{G}(\rho)\} \{2\Phi(c_3) - 1 -2c_3 \Phi'(c_3) \},\\
b_{\rho \rho}(\bm{\Theta}) & = \frac{1}{\sigma^2}\{2\Phi(c_3) - 1\}  \{ \bm{G}(\rho) \bm{Z} \bm{\theta}\}^{\top} \bm{G}(\rho) \bm{Z} \bm{\theta} + 2 \{2\Phi(c_3) - 1 -2c_3 \Phi'(c_3) \} \sum_{i=1}^{n} g_{ii}^2(\rho) \\
 & \ \ \ \ +  \{2\Phi(c_3) - 1\}^2 \sum_{i=1}^{n} \sum_{j\neq i} \{ g_{ij}(\rho) g_{ji}(\rho) + g_{ij}^2(\rho)\},
\end{align*}
where $g_{ij}(\rho)$ is the $(i,j)$\textsuperscript{th} element of $\bm{G}(\rho)$, $\Phi(\cdot)$ and $\Phi'(\cdot)$ are the cdf and pdf of the standard normal distribution, respectively. Therefore, using~\eqref{b_mat}, we get matrix used in~\eqref{if}:
\begin{align*}
	\int_\X \int_\Y \frac{\partial}{\partial {\bm{\Theta}}} \eta_R(\bm{\Theta}; Y)   d F_{\bm{\Theta}}  =  
	\int_\X  \bm{B} (\bm{\Theta}) d F_{\bm{\Theta}}  (\X).
	\end{align*}
A directly calculation yields
\begin{align*}
\int_\X  \int_\Y 	\frac{\partial}{\partial {\rm vec}(\bm{Z})} \eta_{R, \bm{\theta}}(\bm{\Theta}; Y)   \frac{\partial {\rm vec}(\bm{Z})}{\partial \epsilon} \Bigg|_{\epsilon = 0} d F_{\bm{\Theta}} & = \int_\X  \int_\Y (\bm{1}^{\top}_n \bm{\phi}_{2, c_1}(\bm{\epsilon^*})) \bm{R} - \frac{1}{\sigma} \left(\bm{Z}^{\top} \bm{\phi}'_{2, c_1}(\bm{\epsilon^*}) \right) \left(  \bm{\Theta}^{\top}  \bm{R} \right) d F_{\bm{\Theta}} \nonumber\\
& = -  \frac{2\Phi(c_1) - 1}{\sigma}  \int_\X  \left( \bm{Z}^{\top} \bm{1}_n \right) \left( \bm{\Theta}^{\top} \bm{R} \right) d F_{\bm{\Theta}}(\X),
\end{align*}
\begin{equation*}
	\int_\X  \int_\Y 	\frac{\partial}{\partial {\rm vec}(\bm{Z})} \eta_{R, \sigma}(\bm{\Theta}; Y)   \frac{\partial {\rm vec}(\bm{Z})}{\partial \epsilon} \Bigg|_{\epsilon = 0} d F_{\bm{\Theta}}  = -\frac{1}{\sigma} \int_\X  \int_\Y \sum_{i=1}^{j} \sum_{j=1}^{K+1}  \bm{\phi}'_{2, c_2}(\bm{\epsilon^*_i}) \theta_j \left\{ \sum_{l=1}^{n} \bm{\phi}_{2, c_2}(\bm{\epsilon^*_l}) + \bm{\phi}_{2, c_2}(\bm{\epsilon^*_i}) \right\} R_j  d F_{\bm{\Theta}} = 0 ,
\end{equation*}
and
\begin{align*}
	\int_\X  \int_\Y \frac{\partial}{\partial {\rm vec}(\bm{Z})} \eta_{R, \rho}(\bm{\Theta}; Y)   \frac{\partial {\rm vec}(\bm{Z})}{\partial \epsilon} \Bigg|_{\epsilon = 0} d F_{\bm{\Theta}}  &= -\frac{1}{\sigma^2} \int_\X  \int_\Y \sum_{i=1}^{j} \sum_{j=1}^{K+1}  [\bm{G}(\rho) \bm{Z \theta
	}]^{\top} \theta_j \bm{e}_i R_j  d F_{\bm{\Theta}} \\
	&= -\frac{2\Phi(c_3) - 1}{\sigma^2} \int_\X [\bm{G}(\rho) \bm{Z \theta
	}]^{\top} \bm{1}_n (\bm{\theta}^{\top} \bm{R}) d F_{\bm{\Theta}}(\X).
\end{align*}
Here $\bm{1}_n$ is a vector with all elements are one, and the vector $\bm{R}=(R_1, R_2, R_{K+1})^{\top}$ has the first term as $R_1=0$, and the remaining terms form $\text{IF}(\X_0, a, \mathbb{P}_x)$, where $a = (a_1, a_2, \cdots, a_K)^{\top}$. From Theorem 3.1 of \cite{Bali2015}, we find the influence function of the $k$\textsuperscript{th} eigenvalue as:
\begin{equation*}
\text{IF}(\X_0, a_k, \mathbb{P}_x) =  2 a_k \text{IF}\left(\frac{\langle x, \bm{\phi}_k \rangle}{\sqrt{a_k}}  ; \sigma_M, \mathbb{P}_x \right),
\end{equation*}
where $\text{IF}\left(\cdot  ; \sigma_M, \mathbb{P}_x \right)$ is the influence function of the M-scale function for $\sigma_M$:
\begin{equation*}
\text{IF}\left(u  ; \sigma_M, \mathbb{P}_x \right) = \left[ \varphi_{1,c}(u) - \delta\right]  \left\{E_{\mathbb{P}_x}[\varphi_{1,c}'(U)U]\right\}^{-1}, 
\end{equation*}
and the function $\varphi_{1,c}$ is given in~\eqref{eq:loss1}, and $U$ is the standard normal distribution.

As $\bm{\theta}= (\beta_0, \beta^{\top})^{\top}$, we get the influence function of the estimator of $\beta_0$ as the first component of $\text{IF}(\bm{s}, \bm{\Theta}, F_{\bm{\Theta}})$ from~\eqref{if}. The vector containing $2^\textsuperscript{nd}$ to $K^\textsuperscript{th}$ components of~\eqref{if} gives $\text{IF}(\bm{s}, \beta, F_{\bm{\Theta}})$. Finally, we find the influence function of $\widehat{\beta}(t)$ using the relation $\widehat{\beta}(t) = \Xi^{\top} \widehat{\beta}$, $\Xi = (\widehat{\bm{\phi}}_1, \widehat{\bm{\phi}}_2, \cdots, \widehat{\bm{\phi}}_K)^{\top}$, as
\begin{equation}\label{eq:if_b_t}
\text{IF}(\bm{s}, \beta(t), F_\theta) = \Xi^{\top} \text{IF}(\bm{s}, \beta, F_\theta) + \beta^{\top} \text{IF}(\X_0, \Xi, \mathbb{P}_x),
\end{equation}
where $\text{IF}(\X_0, \Xi, \mathbb{P}_x)$ is the influence function of the eigenfunctions. Let 
 $\text{IF}(\X_{k0}, \bm{\phi}_k, \mathbb{P}_x)$ be the $k^\textsuperscript{th}$ the column of $\text{IF}(\X_0, \Xi, \mathbb{P}_x)$ for $k=1,2,\cdots, K$. Then, $\text{IF}(\X_{k0}, \bm{\phi}_k, \mathbb{P}_x)$ is calculated using Theorem 3.1 of \cite{Bali2015}. Let  $\lambda_1 \geq \lambda_2 \geq \cdots$ be the eigenvalues of $\mathcal{C_X}$, the covariance functions of $\X(t)$. If  $\lambda_1 > \lambda_2 > \cdots > \lambda_q > \lambda_{q+1}$, then $\bm{\phi}_j(t)$ are unique up to a sign change when $1\leq j\leq q$. Then, the influence function of the $k$\textsuperscript{th} eigenfunction $\bm{\phi}_k$ for $k\leq q$,  is given by
\begin{equation}
\text{IF}(\X_{0}, \bm{\phi}_k, \mathbb{P}_x) = \sum_{j \geq k +1} \frac{\sqrt{\lambda_k}}{\lambda_k - \lambda_j} \text{DIF}\left( \frac{ \langle \X, \bm{\phi}_k\rangle}{\sqrt{\lambda_k}}, \sigma_M, \mathbb{P}_x\right) \langle \X, \bm{\phi}_j\rangle \bm{\phi}_j  + \sum_{j = 1}^{k-1} \frac{\sqrt{\lambda_j}}{\lambda_k - \lambda_j} \text{DIF}\left( \frac{ \langle \X, \bm{\phi}_j\rangle}{\sqrt{\lambda_j}}, \sigma_M, \mathbb{P}_x \right) \langle \X, \bm{\phi}_k\rangle \bm{\phi}_j, 
\end{equation}
where the derivative of the influence function of $\widehat{\sigma}_M$ is given by
\begin{equation}
\text{DIF}\left( u, \sigma_M, \mathbb{P}_x \right) =  \varphi_{1,c}'(u) \left\{E_{\mathbb{P}_x}[\varphi_{1,c}'(U)U]\right\}^{-1},
\end{equation}
where $U$ is the standard normal distribution.

Note that $\varphi_{0,c}'(u)=0$ when $|u|>c$. So, $\text{DIF}\left( u, \sigma_M, \mathbb{P}_x \right)$ is bounded or even tends to or becomes 0 when $|u|$ converges to $\infty$.  However,  $\text{IF}(\X_{0}, \bm{\phi}_k, \mathbb{P}_x)$ can  be unbounded if there are small scores on some eigenfunctions coupled with large scores on others. Specifically, as highlighted by \cite{Bali2015}, observations with large absolute values of $\langle \X, \bm{\phi}_j\rangle$ and small absolute values of $\langle \X, \bm{\phi}_k\rangle$ for $k < j$ may have a considerable impact on the eigenfunctions. However, $\text{IF}(\bm{s}, \beta, F_\theta)$ is bounded and continuous in $Y$. Therefore, from Equation \eqref{eq:if_b_t}, we conclude that only good leverage points may have an effect on the estimator of $\beta(t)$. Similar results hold for other parameters $\beta_0, \sigma$ and $\rho$.
\end{proof}

\bibliographystyle{agsm}
\bibliography{rsac.bib}

\end{document}